
\catcode`\@=11


\message{Loading jyTeX fonts...}



\font\vptrm=cmr5 \font\vptmit=cmmi5 \font\vptsy=cmsy5 \font\vptbf=cmbx5

\skewchar\vptmit='177 \skewchar\vptsy='60 \fontdimen16
\vptsy=\the\fontdimen17 \vptsy

\def\vpt{\ifmmode\err@badsizechange\else
     \@mathfontinit
     \textfont0=\vptrm  \scriptfont0=\vptrm  \scriptscriptfont0=\vptrm
     \textfont1=\vptmit \scriptfont1=\vptmit \scriptscriptfont1=\vptmit
     \textfont2=\vptsy  \scriptfont2=\vptsy  \scriptscriptfont2=\vptsy
     \textfont3=\xptex  \scriptfont3=\xptex  \scriptscriptfont3=\xptex
     \textfont\bffam=\vptbf
     \scriptfont\bffam=\vptbf
     \scriptscriptfont\bffam=\vptbf
     \@fontstyleinit
     \def\rm{\vptrm\fam=\z@}%
     \def\bf{\vptbf\fam=\bffam}%
     \def\oldstyle{\vptmit\fam=\@ne}%
     \rm\fi}


\font\viptrm=cmr6 \font\viptmit=cmmi6 \font\viptsy=cmsy6
\font\viptbf=cmbx6

\skewchar\viptmit='177 \skewchar\viptsy='60 \fontdimen16
\viptsy=\the\fontdimen17 \viptsy

\def\vipt{\ifmmode\err@badsizechange\else
     \@mathfontinit
     \textfont0=\viptrm  \scriptfont0=\vptrm  \scriptscriptfont0=\vptrm
     \textfont1=\viptmit \scriptfont1=\vptmit \scriptscriptfont1=\vptmit
     \textfont2=\viptsy  \scriptfont2=\vptsy  \scriptscriptfont2=\vptsy
     \textfont3=\xptex   \scriptfont3=\xptex  \scriptscriptfont3=\xptex
     \textfont\bffam=\viptbf
     \scriptfont\bffam=\vptbf
     \scriptscriptfont\bffam=\vptbf
     \@fontstyleinit
     \def\rm{\viptrm\fam=\z@}%
     \def\bf{\viptbf\fam=\bffam}%
     \def\oldstyle{\viptmit\fam=\@ne}%
     \rm\fi}

\font\viiptrm=cmr7 \font\viiptmit=cmmi7 \font\viiptsy=cmsy7
\font\viiptit=cmti7 \font\viiptbf=cmbx7

\skewchar\viiptmit='177 \skewchar\viiptsy='60 \fontdimen16
\viiptsy=\the\fontdimen17 \viiptsy

\def\viipt{\ifmmode\err@badsizechange\else
     \@mathfontinit
     \textfont0=\viiptrm  \scriptfont0=\vptrm  \scriptscriptfont0=\vptrm
     \textfont1=\viiptmit \scriptfont1=\vptmit \scriptscriptfont1=\vptmit
     \textfont2=\viiptsy  \scriptfont2=\vptsy  \scriptscriptfont2=\vptsy
     \textfont3=\xptex    \scriptfont3=\xptex  \scriptscriptfont3=\xptex
     \textfont\itfam=\viiptit
     \scriptfont\itfam=\viiptit
     \scriptscriptfont\itfam=\viiptit
     \textfont\bffam=\viiptbf
     \scriptfont\bffam=\vptbf
     \scriptscriptfont\bffam=\vptbf
     \@fontstyleinit
     \def\rm{\viiptrm\fam=\z@}%
     \def\it{\viiptit\fam=\itfam}%
     \def\bf{\viiptbf\fam=\bffam}%
     \def\oldstyle{\viiptmit\fam=\@ne}%
     \rm\fi}


\font\viiiptrm=cmr8 \font\viiiptmit=cmmi8 \font\viiiptsy=cmsy8
\font\viiiptit=cmti8
\font\viiiptbf=cmbx8

\skewchar\viiiptmit='177 \skewchar\viiiptsy='60 \fontdimen16
\viiiptsy=\the\fontdimen17 \viiiptsy

\def\viiipt{\ifmmode\err@badsizechange\else
     \@mathfontinit
     \textfont0=\viiiptrm  \scriptfont0=\viptrm  \scriptscriptfont0=\vptrm
     \textfont1=\viiiptmit \scriptfont1=\viptmit \scriptscriptfont1=\vptmit
     \textfont2=\viiiptsy  \scriptfont2=\viptsy  \scriptscriptfont2=\vptsy
     \textfont3=\xptex     \scriptfont3=\xptex   \scriptscriptfont3=\xptex
     \textfont\itfam=\viiiptit
     \scriptfont\itfam=\viiptit
     \scriptscriptfont\itfam=\viiptit
     \textfont\bffam=\viiiptbf
     \scriptfont\bffam=\viptbf
     \scriptscriptfont\bffam=\vptbf
     \@fontstyleinit
     \def\rm{\viiiptrm\fam=\z@}%
     \def\it{\viiiptit\fam=\itfam}%
     \def\bf{\viiiptbf\fam=\bffam}%
     \def\oldstyle{\viiiptmit\fam=\@ne}%
     \rm\fi}


\def\getixpt{%
     \font\ixptrm=cmr9
     \font\ixptmit=cmmi9
     \font\ixptsy=cmsy9
     \font\ixptit=cmti9
     \font\ixptbf=cmbx9
     \skewchar\ixptmit='177 \skewchar\ixptsy='60
     \fontdimen16 \ixptsy=\the\fontdimen17 \ixptsy}

\def\ixpt{\ifmmode\err@badsizechange\else
     \@mathfontinit
     \textfont0=\ixptrm  \scriptfont0=\viiptrm  \scriptscriptfont0=\vptrm
     \textfont1=\ixptmit \scriptfont1=\viiptmit \scriptscriptfont1=\vptmit
     \textfont2=\ixptsy  \scriptfont2=\viiptsy  \scriptscriptfont2=\vptsy
     \textfont3=\xptex   \scriptfont3=\xptex    \scriptscriptfont3=\xptex
     \textfont\itfam=\ixptit
     \scriptfont\itfam=\viiptit
     \scriptscriptfont\itfam=\viiptit
     \textfont\bffam=\ixptbf
     \scriptfont\bffam=\viiptbf
     \scriptscriptfont\bffam=\vptbf
     \@fontstyleinit
     \def\rm{\ixptrm\fam=\z@}%
     \def\it{\ixptit\fam=\itfam}%
     \def\bf{\ixptbf\fam=\bffam}%
     \def\oldstyle{\ixptmit\fam=\@ne}%
     \rm\fi}


\font\xptrm=cmr10 \font\xptmit=cmmi10 \font\xptsy=cmsy10
\font\xptex=cmex10 \font\xptit=cmti10 \font\xptsl=cmsl10
\font\xptbf=cmbx10 \font\xpttt=cmtt10 \font\xptss=cmss10
\font\xptsc=cmcsc10 \font\xptbfs=cmb10 \font\xptbmit=cmmib10

\skewchar\xptmit='177 \skewchar\xptbmit='177 \skewchar\xptsy='60
\fontdimen16 \xptsy=\the\fontdimen17 \xptsy

\def\xpt{\ifmmode\err@badsizechange\else
     \@mathfontinit
     \textfont0=\xptrm  \scriptfont0=\viiptrm  \scriptscriptfont0=\vptrm
     \textfont1=\xptmit \scriptfont1=\viiptmit \scriptscriptfont1=\vptmit
     \textfont2=\xptsy  \scriptfont2=\viiptsy  \scriptscriptfont2=\vptsy
     \textfont3=\xptex  \scriptfont3=\xptex    \scriptscriptfont3=\xptex
     \textfont\itfam=\xptit
     \scriptfont\itfam=\viiptit
     \scriptscriptfont\itfam=\viiptit
     \textfont\bffam=\xptbf
     \scriptfont\bffam=\viiptbf
     \scriptscriptfont\bffam=\vptbf
     \textfont\bfsfam=\xptbfs
     \scriptfont\bfsfam=\viiptbf
     \scriptscriptfont\bfsfam=\vptbf
     \textfont\bmitfam=\xptbmit
     \scriptfont\bmitfam=\viiptmit
     \scriptscriptfont\bmitfam=\vptmit
     \@fontstyleinit
     \def\rm{\xptrm\fam=\z@}%
     \def\it{\xptit\fam=\itfam}%
     \def\sl{\xptsl}%
     \def\bf{\xptbf\fam=\bffam}%
     \def\tt{\xpttt}%
     \def\ss{\xptss}%
     \def\sc{\xptsc}%
     \def\bfs{\xptbfs\fam=\bfsfam}%
     \def\bmit{\fam=\bmitfam}%
     \def\oldstyle{\xptmit\fam=\@ne}%
     \rm\fi}


\def\getxipt{%
     \font\xiptrm=cmr10  scaled\magstephalf
     \font\xiptmit=cmmi10 scaled\magstephalf
     \font\xiptsy=cmsy10 scaled\magstephalf
     \font\xiptex=cmex10 scaled\magstephalf
     \font\xiptit=cmti10 scaled\magstephalf
     \font\xiptsl=cmsl10 scaled\magstephalf
     \font\xiptbf=cmbx10 scaled\magstephalf
     \font\xipttt=cmtt10 scaled\magstephalf
     \font\xiptss=cmss10 scaled\magstephalf
     \skewchar\xiptmit='177 \skewchar\xiptsy='60
     \fontdimen16 \xiptsy=\the\fontdimen17 \xiptsy}

\def\xipt{\ifmmode\err@badsizechange\else
     \@mathfontinit
     \textfont0=\xiptrm  \scriptfont0=\viiiptrm  \scriptscriptfont0=\viptrm
     \textfont1=\xiptmit \scriptfont1=\viiiptmit \scriptscriptfont1=\viptmit
     \textfont2=\xiptsy  \scriptfont2=\viiiptsy  \scriptscriptfont2=\viptsy
     \textfont3=\xiptex  \scriptfont3=\xptex     \scriptscriptfont3=\xptex
     \textfont\itfam=\xiptit
     \scriptfont\itfam=\viiiptit
     \scriptscriptfont\itfam=\viiptit
     \textfont\bffam=\xiptbf
     \scriptfont\bffam=\viiiptbf
     \scriptscriptfont\bffam=\viptbf
     \@fontstyleinit
     \def\rm{\xiptrm\fam=\z@}%
     \def\it{\xiptit\fam=\itfam}%
     \def\sl{\xiptsl}%
     \def\bf{\xiptbf\fam=\bffam}%
     \def\tt{\xipttt}%
     \def\ss{\xiptss}%
     \def\oldstyle{\xiptmit\fam=\@ne}%
     \rm\fi}


\font\xiiptrm=cmr12 \font\xiiptmit=cmmi12 \font\xiiptsy=cmsy10
scaled\magstep1 \font\xiiptex=cmex10  scaled\magstep1
\font\xiiptit=cmti12 \font\xiiptsl=cmsl12 \font\xiiptbf=cmbx12
\font\xiiptss=cmss12 \font\xiiptsc=cmcsc10 scaled\magstep1
\font\xiiptbfs=cmb10  scaled\magstep1 \font\xiiptbmit=cmmib10
scaled\magstep1

\skewchar\xiiptmit='177 \skewchar\xiiptbmit='177 \skewchar\xiiptsy='60
\fontdimen16 \xiiptsy=\the\fontdimen17 \xiiptsy

\def\xiipt{\ifmmode\err@badsizechange\else
     \@mathfontinit
     \textfont0=\xiiptrm  \scriptfont0=\viiiptrm  \scriptscriptfont0=\viptrm
     \textfont1=\xiiptmit \scriptfont1=\viiiptmit \scriptscriptfont1=\viptmit
     \textfont2=\xiiptsy  \scriptfont2=\viiiptsy  \scriptscriptfont2=\viptsy
     \textfont3=\xiiptex  \scriptfont3=\xptex     \scriptscriptfont3=\xptex
     \textfont\itfam=\xiiptit
     \scriptfont\itfam=\viiiptit
     \scriptscriptfont\itfam=\viiptit
     \textfont\bffam=\xiiptbf
     \scriptfont\bffam=\viiiptbf
     \scriptscriptfont\bffam=\viptbf
     \textfont\bfsfam=\xiiptbfs
     \scriptfont\bfsfam=\viiiptbf
     \scriptscriptfont\bfsfam=\viptbf
     \textfont\bmitfam=\xiiptbmit
     \scriptfont\bmitfam=\viiiptmit
     \scriptscriptfont\bmitfam=\viptmit
     \@fontstyleinit
     \def\rm{\xiiptrm\fam=\z@}%
     \def\it{\xiiptit\fam=\itfam}%
     \def\sl{\xiiptsl}%
     \def\bf{\xiiptbf\fam=\bffam}%
     \def\tt{\xiipttt}%
     \def\ss{\xiiptss}%
     \def\sc{\xiiptsc}%
     \def\bfs{\xiiptbfs\fam=\bfsfam}%
     \def\bmit{\fam=\bmitfam}%
     \def\oldstyle{\xiiptmit\fam=\@ne}%
     \rm\fi}


\def\getxiiipt{%
     \font\xiiiptrm=cmr12  scaled\magstephalf
     \font\xiiiptmit=cmmi12 scaled\magstephalf
     \font\xiiiptsy=cmsy9  scaled\magstep2
     \font\xiiiptit=cmti12 scaled\magstephalf
     \font\xiiiptsl=cmsl12 scaled\magstephalf
     \font\xiiiptbf=cmbx12 scaled\magstephalf
     \font\xiiipttt=cmtt12 scaled\magstephalf
     \font\xiiiptss=cmss12 scaled\magstephalf
     \skewchar\xiiiptmit='177 \skewchar\xiiiptsy='60
     \fontdimen16 \xiiiptsy=\the\fontdimen17 \xiiiptsy}

\def\xiiipt{\ifmmode\err@badsizechange\else
     \@mathfontinit
     \textfont0=\xiiiptrm  \scriptfont0=\xptrm  \scriptscriptfont0=\viiptrm
     \textfont1=\xiiiptmit \scriptfont1=\xptmit \scriptscriptfont1=\viiptmit
     \textfont2=\xiiiptsy  \scriptfont2=\xptsy  \scriptscriptfont2=\viiptsy
     \textfont3=\xivptex   \scriptfont3=\xptex  \scriptscriptfont3=\xptex
     \textfont\itfam=\xiiiptit
     \scriptfont\itfam=\xptit
     \scriptscriptfont\itfam=\viiptit
     \textfont\bffam=\xiiiptbf
     \scriptfont\bffam=\xptbf
     \scriptscriptfont\bffam=\viiptbf
     \@fontstyleinit
     \def\rm{\xiiiptrm\fam=\z@}%
     \def\it{\xiiiptit\fam=\itfam}%
     \def\sl{\xiiiptsl}%
     \def\bf{\xiiiptbf\fam=\bffam}%
     \def\tt{\xiiipttt}%
     \def\ss{\xiiiptss}%
     \def\oldstyle{\xiiiptmit\fam=\@ne}%
     \rm\fi}


\font\xivptrm=cmr12   scaled\magstep1 \font\xivptmit=cmmi12
scaled\magstep1 \font\xivptsy=cmsy10  scaled\magstep2
\font\xivptex=cmex10  scaled\magstep2 \font\xivptit=cmti12
scaled\magstep1 \font\xivptsl=cmsl12  scaled\magstep1
\font\xivptbf=cmbx12  scaled\magstep1
\font\xivptss=cmss12  scaled\magstep1 \font\xivptsc=cmcsc10
scaled\magstep2 \font\xivptbfs=cmb10  scaled\magstep2
\font\xivptbmit=cmmib10 scaled\magstep2

\skewchar\xivptmit='177 \skewchar\xivptbmit='177 \skewchar\xivptsy='60
\fontdimen16 \xivptsy=\the\fontdimen17 \xivptsy

\def\xivpt{\ifmmode\err@badsizechange\else
     \@mathfontinit
     \textfont0=\xivptrm  \scriptfont0=\xptrm  \scriptscriptfont0=\viiptrm
     \textfont1=\xivptmit \scriptfont1=\xptmit \scriptscriptfont1=\viiptmit
     \textfont2=\xivptsy  \scriptfont2=\xptsy  \scriptscriptfont2=\viiptsy
     \textfont3=\xivptex  \scriptfont3=\xptex  \scriptscriptfont3=\xptex
     \textfont\itfam=\xivptit
     \scriptfont\itfam=\xptit
     \scriptscriptfont\itfam=\viiptit
     \textfont\bffam=\xivptbf
     \scriptfont\bffam=\xptbf
     \scriptscriptfont\bffam=\viiptbf
     \textfont\bfsfam=\xivptbfs
     \scriptfont\bfsfam=\xptbfs
     \scriptscriptfont\bfsfam=\viiptbf
     \textfont\bmitfam=\xivptbmit
     \scriptfont\bmitfam=\xptbmit
     \scriptscriptfont\bmitfam=\viiptmit
     \@fontstyleinit
     \def\rm{\xivptrm\fam=\z@}%
     \def\it{\xivptit\fam=\itfam}%
     \def\sl{\xivptsl}%
     \def\bf{\xivptbf\fam=\bffam}%
     \def\tt{\xivpttt}%
     \def\ss{\xivptss}%
     \def\sc{\xivptsc}%
     \def\bfs{\xivptbfs\fam=\bfsfam}%
     \def\bmit{\fam=\bmitfam}%
     \def\oldstyle{\xivptmit\fam=\@ne}%
     \rm\fi}


\font\xviiptrm=cmr17 \font\xviiptmit=cmmi12 scaled\magstep2
\font\xviiptsy=cmsy10 scaled\magstep3 \font\xviiptex=cmex10
scaled\magstep3 \font\xviiptit=cmti12 scaled\magstep2
\font\xviiptbf=cmbx12 scaled\magstep2 \font\xviiptbfs=cmb10
scaled\magstep3

\skewchar\xviiptmit='177 \skewchar\xviiptsy='60 \fontdimen16
\xviiptsy=\the\fontdimen17 \xviiptsy

\def\xviipt{\ifmmode\err@badsizechange\else
     \@mathfontinit
     \textfont0=\xviiptrm  \scriptfont0=\xiiptrm  \scriptscriptfont0=\viiiptrm
     \textfont1=\xviiptmit \scriptfont1=\xiiptmit \scriptscriptfont1=\viiiptmit
     \textfont2=\xviiptsy  \scriptfont2=\xiiptsy  \scriptscriptfont2=\viiiptsy
     \textfont3=\xviiptex  \scriptfont3=\xiiptex  \scriptscriptfont3=\xptex
     \textfont\itfam=\xviiptit
     \scriptfont\itfam=\xiiptit
     \scriptscriptfont\itfam=\viiiptit
     \textfont\bffam=\xviiptbf
     \scriptfont\bffam=\xiiptbf
     \scriptscriptfont\bffam=\viiiptbf
     \textfont\bfsfam=\xviiptbfs
     \scriptfont\bfsfam=\xiiptbfs
     \scriptscriptfont\bfsfam=\viiiptbf
     \@fontstyleinit
     \def\rm{\xviiptrm\fam=\z@}%
     \def\it{\xviiptit\fam=\itfam}%
     \def\bf{\xviiptbf\fam=\bffam}%
     \def\bfs{\xviiptbfs\fam=\bfsfam}%
     \def\oldstyle{\xviiptmit\fam=\@ne}%
     \rm\fi}


\font\xxiptrm=cmr17  scaled\magstep1


\def\xxipt{\ifmmode\err@badsizechange\else
     \@mathfontinit
     \@fontstyleinit
     \def\rm{\xxiptrm\fam=\z@}%
     \rm\fi}


\font\xxvptrm=cmr17  scaled\magstep2


\def\xxvpt{\ifmmode\err@badsizechange\else
     \@mathfontinit
     \@fontstyleinit
     \def\rm{\xxvptrm\fam=\z@}%
     \rm\fi}




\message{Loading jyTeX macros...}

\message{modifications to plain.tex,}


\def\newcount{\alloc@0\count\countdef\insc@unt}
\def\newdimen{\alloc@1\dimen\dimendef\insc@unt}
\def\newskip{\alloc@2\skip\skipdef\insc@unt}
\def\newmuskip{\alloc@3\muskip\muskipdef\@cclvi}
\def\newbox{\alloc@4\box\chardef\insc@unt}
\def\newtoks{\alloc@5\toks\toksdef\@cclvi}
\def\newhelp#1#2{\newtoks#1\global#1\expandafter{\csname#2\endcsname}}
\def\newread{\alloc@6\read\chardef\sixt@@n}
\def\newwrite{\alloc@7\write\chardef\sixt@@n}
\def\newfam{\alloc@8\fam\chardef\sixt@@n}
\def\newinsert#1{\global\advance\insc@unt by\m@ne
     \ch@ck0\insc@unt\count
     \ch@ck1\insc@unt\dimen
     \ch@ck2\insc@unt\skip
     \ch@ck4\insc@unt\box
     \allocationnumber=\insc@unt
     \global\chardef#1=\allocationnumber
     \wlog{\string#1=\string\insert\the\allocationnumber}}
\def\newif#1{\count@\escapechar \escapechar\m@ne
     \expandafter\expandafter\expandafter
          \xdef\@if#1{true}{\let\noexpand#1=\noexpand\iftrue}%
     \expandafter\expandafter\expandafter
          \xdef\@if#1{false}{\let\noexpand#1=\noexpand\iffalse}%
     \global\@if#1{false}\escapechar=\count@}


\newlinechar=`\^^J
\overfullrule=0pt




\let\itfam=\undefined

\let\bffam=\undefined

\count18=3


\chardef\sharps="19


\mathchardef\alpha="710B \mathchardef\beta="710C \mathchardef\gamma="710D
\mathchardef\delta="710E \mathchardef\epsilon="710F
\mathchardef\zeta="7110 \mathchardef\eta="7111 \mathchardef\theta="7112
\mathchardef\iota="7113 \mathchardef\kappa="7114
\mathchardef\lambda="7115 \mathchardef\mu="7116 \mathchardef\nu="7117
\mathchardef\xi="7118 \mathchardef\pi="7119 \mathchardef\rho="711A
\mathchardef\sigma="711B \mathchardef\tau="711C
\mathchardef\upsilon="711D \mathchardef\phi="711E \mathchardef\chi="711F
\mathchardef\psi="7120 \mathchardef\omega="7121
\mathchardef\varepsilon="7122 \mathchardef\vartheta="7123
\mathchardef\varpi="7124 \mathchardef\varrho="7125
\mathchardef\varsigma="7126 \mathchardef\varphi="7127
\mathchardef\imath="717B \mathchardef\jmath="717C \mathchardef\ell="7160
\mathchardef\wp="717D \mathchardef\partial="7140 \mathchardef\flat="715B
\mathchardef\natural="715C \mathchardef\sharp="715D



\def\angle{{\vbox{\ialign{$\m@th\scriptstyle##$\crcr
     \not\mathrel{\mkern14mu}\crcr
     \noalign{\nointerlineskip}
     \mkern2.5mu\leaders\hrule height.34\rp@\hfill\mkern2.5mu\crcr}}}}
\def\vdots{\vbox{\baselineskip4\rp@ \lineskiplimit\z@
     \kern6\rp@\hbox{.}\hbox{.}\hbox{.}}}
\def\ddots{\mathinner{\mkern1mu\raise7\rp@\vbox{\kern7\rp@\hbox{.}}\mkern2mu
     \raise4\rp@\hbox{.}\mkern2mu\raise\rp@\hbox{.}\mkern1mu}}
\def\overrightarrow#1{\vbox{\ialign{##\crcr
     \rightarrowfill\crcr
     \noalign{\kern-\rp@\nointerlineskip}
     $\hfil\displaystyle{#1}\hfil$\crcr}}}
\def\overleftarrow#1{\vbox{\ialign{##\crcr
     \leftarrowfill\crcr
     \noalign{\kern-\rp@\nointerlineskip}
     $\hfil\displaystyle{#1}\hfil$\crcr}}}
\def\overbrace#1{\mathop{\vbox{\ialign{##\crcr
     \noalign{\kern3\rp@}
     \downbracefill\crcr
     \noalign{\kern3\rp@\nointerlineskip}
     $\hfil\displaystyle{#1}\hfil$\crcr}}}\limits}
\def\underbrace#1{\mathop{\vtop{\ialign{##\crcr
     $\hfil\displaystyle{#1}\hfil$\crcr
     \noalign{\kern3\rp@\nointerlineskip}
     \upbracefill\crcr
     \noalign{\kern3\rp@}}}}\limits}
\def\big#1{{\hbox{$\left#1\vbox to8.5\rp@ {}\right.\n@space$}}}
\def\Big#1{{\hbox{$\left#1\vbox to11.5\rp@ {}\right.\n@space$}}}
\def\bigg#1{{\hbox{$\left#1\vbox to14.5\rp@ {}\right.\n@space$}}}
\def\Bigg#1{{\hbox{$\left#1\vbox to17.5\rp@ {}\right.\n@space$}}}
\def\@vereq#1#2{\lower.5\rp@\vbox{\baselineskip\z@skip\lineskip-.5\rp@
     \ialign{$\m@th#1\hfil##\hfil$\crcr#2\crcr=\crcr}}}
\def\rlh@#1{\vcenter{\hbox{\ooalign{\raise2\rp@
     \hbox{$#1\rightharpoonup$}\crcr
     $#1\leftharpoondown$}}}}
\def\bordermatrix#1{\begingroup\m@th
     \setbox\z@\vbox{%
          \def\cr{\crcr\noalign{\kern2\rp@\global\let\cr\endline}}%
          \ialign{$##$\hfil\kern2\rp@\kern\p@renwd
               &\thinspace\hfil$##$\hfil&&\quad\hfil$##$\hfil\crcr
               \omit\strut\hfil\crcr
               \noalign{\kern-\baselineskip}%
               #1\crcr\omit\strut\cr}}%
     \setbox\tw@\vbox{\unvcopy\z@\global\setbox\@ne\lastbox}%
     \setbox\tw@\hbox{\unhbox\@ne\unskip\global\setbox\@ne\lastbox}%
     \setbox\tw@\hbox{$\kern\wd\@ne\kern-\p@renwd\left(\kern-\wd\@ne
          \global\setbox\@ne\vbox{\box\@ne\kern2\rp@}%
          \vcenter{\kern-\ht\@ne\unvbox\z@\kern-\baselineskip}%
          \,\right)$}%
     \null\;\vbox{\kern\ht\@ne\box\tw@}\endgroup}
\def\endinsert{\egroup
     \if@mid\dimen@\ht\z@
          \advance\dimen@\dp\z@
          \advance\dimen@12\rp@
          \advance\dimen@\pagetotal
          \ifdim\dimen@>\pagegoal\@midfalse\p@gefalse\fi
     \fi
     \if@mid\bigskip\box\z@
          \bigbreak
     \else\insert\topins{\penalty100 \splittopskip\z@skip
               \splitmaxdepth\maxdimen\floatingpenalty\z@
               \ifp@ge\dimen@\dp\z@
                    \vbox to\vsize{\unvbox\z@\kern-\dimen@}%
               \else\box\z@\nobreak\bigskip
               \fi}%
     \fi
     \endgroup}


\def\cases#1{\left\{\,\vcenter{\m@th
     \ialign{$##\hfil$&\quad##\hfil\crcr#1\crcr}}\right.}
\def\matrix#1{\null\,\vcenter{\m@th
     \ialign{\hfil$##$\hfil&&\quad\hfil$##$\hfil\crcr
          \mathstrut\crcr
          \noalign{\kern-\baselineskip}
          #1\crcr
          \mathstrut\crcr
          \noalign{\kern-\baselineskip}}}\,}


\newif\ifraggedbottom

\def\raggedbottom{\ifraggedbottom\else
     \advance\topskip by\z@ plus60pt \raggedbottomtrue\fi}%
\def\normalbottom{\ifraggedbottom
     \advance\topskip by\z@ plus-60pt \raggedbottomfalse\fi}

\message{hacks,}


\toksdef\toks@i=1 \toksdef\toks@ii=2


\def\TeX{T\kern-.1667em \lower.5ex \hbox{E}\kern-.125em X\null}
\def\jyTeX{{\leavevmode
     \raise.587ex \hbox{\it\j}\kern-.1em \lower.048ex \hbox{\it y}\kern-.12em
     \TeX}}

\let\then=\iftrue
\def\ifnoarg#1\then{\def\hack@{#1}\ifx\hack@\empty}
\def\ifundefined#1\then{%
     \expandafter\ifx\csname\expandafter\blank\string#1\endcsname\relax}
\def\useif#1\then{\csname#1\endcsname}
\def\usename#1{\csname#1\endcsname}
\def\useafter#1#2{\expandafter#1\csname#2\endcsname}

\long\def\loop#1\repeat{\def\@iterate{#1\expandafter\@iterate\fi}\@iterate
     \let\@iterate=\relax}

\let\TeXend=\end
\def\begin#1{\begingroup\def\@@blockname{#1}\usename{begin#1}}
\def\end#1{\usename{end#1}\def\hack@{#1}%
     \ifx\@@blockname\hack@
          \endgroup
     \else\err@badgroup\hack@\@@blockname
     \fi}
\def\@@blockname{}

\def\defaultoption[#1]#2{%
     \def\hack@{\ifx\hack@ii[\toks@={#2}\else\toks@={#2[#1]}\fi\the\toks@}%
     \futurelet\hack@ii\hack@}

\def\markup#1{\let\@@marksf=\empty
     \ifhmode\edef\@@marksf{\spacefactor=\the\spacefactor\relax}\/\fi
     ${}^{\hbox{\subscriptfonts#1}}$\@@marksf}


\newtoks\shortyear
\newtoks\militaryhour
\newtoks\standardhour
\newtoks\minute
\newtoks\amorpm

\def\settime{\count@=\time\divide\count@ by60
     \militaryhour=\expandafter{\number\count@}%
     {\multiply\count@ by-60 \advance\count@ by\time
          \xdef\hack@{\ifnum\count@<10 0\fi\number\count@}}%
     \minute=\expandafter{\hack@}%
     \ifnum\count@<12
          \amorpm={am}
     \else\amorpm={pm}
          \ifnum\count@>12 \advance\count@ by-12 \fi
     \fi
     \standardhour=\expandafter{\number\count@}%
     \def\hack@19##1##2{\shortyear={##1##2}}%
          \expandafter\hack@\the\year}

\def\monthword#1{%
     \ifcase#1
          $\bullet$\err@badcountervalue{monthword}%
          \or January\or February\or March\or April\or May\or June%
          \or July\or August\or September\or October\or November\or December%
     \else$\bullet$\err@badcountervalue{monthword}%
     \fi}

\def\monthabbr#1{%
     \ifcase#1
          $\bullet$\err@badcountervalue{monthabbr}%
          \or Jan\or Feb\or Mar\or Apr\or May\or Jun%
          \or Jul\or Aug\or Sep\or Oct\or Nov\or Dec%
     \else$\bullet$\err@badcountervalue{monthabbr}%
     \fi}

\def\militarytime{\the\militaryhour:\the\minute}
\def\standardtime{\the\standardhour:\the\minute}


\def\@setnumstyle#1#2{\expandafter\global\expandafter\expandafter
     \expandafter\let\expandafter\expandafter
     \csname @\expandafter\blank\string#1style\endcsname
     \csname#2\endcsname}
\def\numstyle#1{\usename{@\expandafter\blank\string#1style}#1}
\def\ifblank#1\then{\useafter\ifx{@\expandafter\blank\string#1}\blank}

\def\blank#1{}

\def\Roman#1{\expandafter\uppercase\expandafter{\romannumeral#1}}
\def\alphabetic#1{%
     \ifcase#1
          $\bullet$\err@badcountervalue{alphabetic}%
          \or a\or b\or c\or d\or e\or f\or g\or h\or i\or j\or k\or l\or m%
          \or n\or o\or p\or q\or r\or s\or t\or u\or v\or w\or x\or y\or z%
     \else$\bullet$\err@badcountervalue{alphabetic}%
     \fi}
\def\Alphabetic#1{\expandafter\uppercase\expandafter{\alphabetic{#1}}}
\def\symbols#1{%
     \ifcase#1
          $\bullet$\err@badcountervalue{symbols}%
          \or*\or\dag\or\ddag\or\S\or$\|$%
          \or**\or\dag\dag\or\ddag\ddag\or\S\S\or$\|\|$%
     \else$\bullet$\err@badcountervalue{symbols}%
     \fi}


\catcode`\^^?=13 \def^^?{\relax}

\def\trimleading#1\to#2{\edef#2{#1}%
     \expandafter\@trimleading\expandafter#2#2^^?^^?}
\def\@trimleading#1#2#3^^?{\ifx#2^^?\def#1{}\else\def#1{#2#3}\fi}

\def\trimtrailing#1\to#2{\edef#2{#1}%
     \expandafter\@trimtrailing\expandafter#2#2^^? ^^?\relax}
\def\@trimtrailing#1#2 ^^?#3{\ifx#3\relax\toks@={}%
     \else\def#1{#2}\toks@={\trimtrailing#1\to#1}\fi
     \the\toks@}

\def\trim#1\to#2{\trimleading#1\to#2\trimtrailing#2\to#2}

\catcode`\^^?=15


\long\def\additemL#1\to#2{\toks@={\^^\{#1}}\toks@ii=\expandafter{#2}%
     \xdef#2{\the\toks@\the\toks@ii}}

\long\def\additemR#1\to#2{\toks@={\^^\{#1}}\toks@ii=\expandafter{#2}%
     \xdef#2{\the\toks@ii\the\toks@}}

\def\getitemL#1\to#2{\expandafter\@getitemL#1\hack@#1#2}
\def\@getitemL\^^\#1#2\hack@#3#4{\def#4{#1}\def#3{#2}}

\message{font macros,}


\newdimen\rp@
\newcount\@@sizeindex \@@sizeindex=0
\newcount\@@factori
\newcount\@@factorii
\newcount\@@factoriii
\newcount\@@factoriv

\countdef\maxfam=18
\newfam\itfam
\newfam\bffam
\newfam\bfsfam
\newfam\bmitfam

\def\@mathfontinit{\count@=4
     \loop\textfont\count@=\nullfont
          \scriptfont\count@=\nullfont
          \scriptscriptfont\count@=\nullfont
          \ifnum\count@<\maxfam\advance\count@ by\@ne
     \repeat}

\def\@fontstyleinit{%
     \def\it{\err@fontnotavailable\it}%
     \def\bf{\err@fontnotavailable\bf}%
     \def\bfs{\err@bfstobf}%
     \def\bmit{\err@fontnotavailable\bmit}%
     \def\sc{\err@fontnotavailable\sc}%
     \def\sl{\err@sltoit}%
     \def\ss{\err@fontnotavailable\ss}%
     \def\tt{\err@fontnotavailable\tt}}

\def\@parameterinit#1{\rm\rp@=.1em \@getscaling{#1}%
     \let\^^\=\@doscaling\scalingskipslist
     \setbox\strutbox=\hbox{\vrule
          height.708\baselineskip depth.292\baselineskip width\z@}}

\def\@getfactor#1#2#3#4{\@@factori=#1 \@@factorii=#2
     \@@factoriii=#3 \@@factoriv=#4}

\def\@getscaling#1{\count@=#1 \advance\count@ by-\@@sizeindex\@@sizeindex=#1
     \ifnum\count@<0
          \let\@mulordiv=\divide
          \let\@divormul=\multiply
          \multiply\count@ by\m@ne
     \else\let\@mulordiv=\multiply
          \let\@divormul=\divide
     \fi
     \edef\@@scratcha{\ifcase\count@                {1}{1}{1}{1}\or
          {1}{7}{23}{3}\or     {2}{5}{3}{1}\or      {9}{89}{13}{1}\or
          {6}{25}{6}{1}\or     {8}{71}{14}{1}\or    {6}{25}{36}{5}\or
          {1}{7}{53}{4}\or     {12}{125}{108}{5}\or {3}{14}{53}{5}\or
          {6}{41}{17}{1}\or    {13}{31}{13}{2}\or   {9}{107}{71}{2}\or
          {11}{139}{124}{3}\or {1}{6}{43}{2}\or     {10}{107}{42}{1}\or
          {1}{5}{43}{2}\or     {5}{69}{65}{1}\or    {11}{97}{91}{2}\fi}%
     \expandafter\@getfactor\@@scratcha}

\def\@doscaling#1{\@mulordiv#1by\@@factori\@divormul#1by\@@factorii
     \@mulordiv#1by\@@factoriii\@divormul#1by\@@factoriv}


\newskip\headskip
\newskip\footskip

\def\typesize=#1pt{\count@=#1 \advance\count@ by-10
     \ifcase\count@
          \@setsizex\or\err@badtypesize\or
          \@setsizexii\or\err@badtypesize\or
          \@setsizexiv
     \else\err@badtypesize
     \fi}

\def\@setsizex{\getixpt
     \def\subsubscriptfonts{\vpt}%
          \def\subsubscriptsize{\vpt\@parameterinit{-8}}%
     \def\subscriptfonts{\viipt}\def\subscriptsize{\viipt\@parameterinit{-4}}%
     \def\footnotefonts{\viiipt}\def\footnotesize{\viiipt\@parameterinit{-2}}%
     \def\smallfonts{\ixpt}\def\smallsize{\ixpt\@parameterinit{-1}}%
     \def\normalfonts{\xpt}\def\normalsize{\xpt\@parameterinit{0}}%
     \def\bigfonts{\xiipt}\def\bigsize{\xiipt\@parameterinit{2}}%
     \def\Bigfonts{\xivpt}\def\Bigsize{\xivpt\@parameterinit{4}}%
     \def\biggfonts{\xviipt}\def\biggsize{\xviipt\@parameterinit{6}}%
     \def\Biggfonts{\xxipt}\def\Biggsize{\xxipt\@parameterinit{8}}%
     \def\tinyfonts{\vpt}\def\tinysize{\vpt\@parameterinit{-8}}%
     \def\HUGEFONTS{\xxvpt}\def\HUGESIZE{\xxvpt\@parameterinit{10}}%
     \normalsize\fixedskipslist}

\def\@setsizexii{\getxipt
     \def\subsubscriptfonts{\vipt}%
          \def\subsubscriptsize{\vipt\@parameterinit{-6}}%
     \def\subscriptfonts{\viiipt}%
          \def\subscriptsize{\viiipt\@parameterinit{-2}}%
     \def\footnotefonts{\xpt}\def\footnotesize{\xpt\@parameterinit{0}}%
     \def\smallfonts{\xipt}\def\smallsize{\xipt\@parameterinit{1}}%
     \def\normalfonts{\xiipt}\def\normalsize{\xiipt\@parameterinit{2}}%
     \def\bigfonts{\xivpt}\def\bigsize{\xivpt\@parameterinit{4}}%
     \def\Bigfonts{\xviipt}\def\Bigsize{\xviipt\@parameterinit{6}}%
     \def\biggfonts{\xxipt}\def\biggsize{\xxipt\@parameterinit{8}}%
     \def\Biggfonts{\xxvpt}\def\Biggsize{\xxvpt\@parameterinit{10}}%
     \def\tinyfonts{\vpt}\def\tinysize{\vpt\@parameterinit{-8}}%
     \def\HUGEFONTS{\xxvpt}\def\HUGESIZE{\xxvpt\@parameterinit{10}}%
     \normalsize\fixedskipslist}

\def\@setsizexiv{\getxiiipt
     \def\subsubscriptfonts{\viipt}%
          \def\subsubscriptsize{\viipt\@parameterinit{-4}}%
     \def\subscriptfonts{\xpt}\def\subscriptsize{\xpt\@parameterinit{0}}%
     \def\footnotefonts{\xiipt}\def\footnotesize{\xiipt\@parameterinit{2}}%
     \def\smallfonts{\xiiipt}\def\smallsize{\xiiipt\@parameterinit{3}}%
     \def\normalfonts{\xivpt}\def\normalsize{\xivpt\@parameterinit{4}}%
     \def\bigfonts{\xviipt}\def\bigsize{\xviipt\@parameterinit{6}}%
     \def\Bigfonts{\xxipt}\def\Bigsize{\xxipt\@parameterinit{8}}%
     \def\biggfonts{\xxvpt}\def\biggsize{\xxvpt\@parameterinit{10}}%
     \def\Biggfonts{\err@sizetoolarge\Biggfonts\HUGEFONTS}%
          \def\Biggsize{\err@sizetoolarge\Biggsize\HUGESIZE}%
     \def\tinyfonts{\vpt}\def\tinysize{\vpt\@parameterinit{-8}}%
     \def\HUGEFONTS{\xxvpt}\def\HUGESIZE{\xxvpt\@parameterinit{10}}%
     \normalsize\fixedskipslist}

\def\subsubscriptfonts{\vpt} \def\subsubscriptsize{\vpt\@parameterinit{-8}}
\def\subscriptfonts{\viipt}  \def\subscriptsize{\viipt\@parameterinit{-4}}
\def\footnotefonts{\viiipt}  \def\footnotesize{\viiipt\@parameterinit{-2}}
\def\smallfonts{\err@sizenotavailable\smallfonts}
                             \def\smallsize{\ixpt\@parameterinit{-1}}
\def\normalfonts{\xpt}       \def\normalsize{\xpt\@parameterinit{0}}
\def\bigfonts{\xiipt}        \def\bigsize{\xiipt\@parameterinit{2}}
\def\Bigfonts{\xivpt}        \def\Bigsize{\xivpt\@parameterinit{4}}
\def\biggfonts{\xviipt}      \def\biggsize{\xviipt\@parameterinit{6}}
\def\Biggfonts{\xxipt}       \def\Biggsize{\xxipt\@parameterinit{8}}
\def\tinyfonts{\vpt}         \def\tinysize{\vpt\@parameterinit{-8}}
\def\HUGEFONTS{\xxvpt}       \def\HUGESIZE{\xxvpt\@parameterinit{10}}

\message{document layout,}


\newtoks\everyoutput \everyoutput={}
\newdimen\depthofpage
\newcount\pagenum \pagenum=0

\newdimen\oddtopmargin  \newdimen\eventopmargin
\newdimen\oddleftmargin \newdimen\evenleftmargin
\newtoks\oddhead        \newtoks\evenhead
\newtoks\oddfoot        \newtoks\evenfoot

\def\topmargin{\afterassignment\@seteventop\oddtopmargin}
\def\leftmargin{\afterassignment\@setevenleft\oddleftmargin}
\def\head{\afterassignment\@setevenhead\oddhead}
\def\foot{\afterassignment\@setevenfoot\oddfoot}

\def\@seteventop{\eventopmargin=\oddtopmargin}
\def\@setevenleft{\evenleftmargin=\oddleftmargin}
\def\@setevenhead{\evenhead=\oddhead}
\def\@setevenfoot{\evenfoot=\oddfoot}

\def\pagenumstyle#1{\@setnumstyle\pagenum{#1}}

\newif\ifdraft
\def\draft{\drafttrue\leftmargin=.5in \overfullrule=5pt }

\def\outputstyle#1{\global\expandafter\let\expandafter
          \@outputstyle\csname#1output\endcsname
     \usename{#1setup}}

\output={\@outputstyle}

\def\normaloutput{\the\everyoutput
     \global\advance\pagenum by\@ne
     \ifodd\pagenum
          \voffset=\oddtopmargin \hoffset=\oddleftmargin
     \else\voffset=\eventopmargin \hoffset=\evenleftmargin
     \fi
     \advance\voffset by-1in  \advance\hoffset by-1in
     \count0=\pagenum
     \expandafter\shipout\pagebox
     \ifnum\outputpenalty>-\@MM\else\dosupereject\fi}

\newdimen\fullhsize
\newbox\leftpage
\newcount\leftpagenum
\newcount\outputpagenum \outputpagenum=0
\let\leftorright=L

\def\twoupoutput{\the\everyoutput
     \global\advance\pagenum by\@ne
     \if L\leftorright
          \global\setbox\leftpage=\leftline{\pagebox}%
          \global\leftpagenum=\pagenum
          \global\let\leftorright=R%
     \else\global\advance\outputpagenum by\@ne
          \ifodd\outputpagenum
               \voffset=\oddtopmargin \hoffset=\oddleftmargin
          \else\voffset=\eventopmargin \hoffset=\evenleftmargin
          \fi
          \advance\voffset by-1in  \advance\hoffset by-1in
          \count0=\leftpagenum \count1=\pagenum
          \shipout\vbox{\hbox to\fullhsize
               {\box\leftpage\hfil\leftline{\pagebox}}}%
          \global\let\leftorright=L%
     \fi
     \ifnum\outputpenalty>-\@MM
     \else\dosupereject
          \if R\leftorright
               \globaldefs=\@ne\head={\hfil}\foot={\hfil}\globaldefs=\z@
               \null\newpage
          \fi
     \fi}

\def\pagebox{\vbox{\makeheadline\pagebody\makefootline}}

\def\makeheadline{%
     \vbox to\z@{\baselinestretch=\@m
          \vskip\topskip\vskip-.708\baselineskip\vskip-\headskip
          \line{\vbox to\ht\strutbox{}%
               \ifodd\pagenum\the\oddhead\else\the\evenhead\fi}%
          \vss}%
     \nointerlineskip}

\def\pagebody{\vbox to\vsize{%
     \boxmaxdepth\maxdepth
     \ifvoid\topins\else\unvbox\topins\fi
     \depthofpage=\dp255
     \unvbox255
     \ifraggedbottom\kern-\depthofpage\vfil\fi
     \ifvoid\footins
     \else\vskip\skip\footins
          \footnoterule
          \unvbox\footins
          \vskip-\footnoteskip
     \fi}}

\def\makefootline{\baselineskip=\footskip
     \line{\ifodd\pagenum\the\oddfoot\else\the\evenfoot\fi}}


\newskip\abovechapterskip
\newskip\belowchapterskip
\newskip\abovesectionskip
\newskip\belowsectionskip
\newskip\abovesubsectionskip
\newskip\belowsubsectionskip

\def\chapterstyle#1{\global\expandafter\let\expandafter\@chapterstyle
     \csname#1text\endcsname}
\def\sectionstyle#1{\global\expandafter\let\expandafter\@sectionstyle
     \csname#1text\endcsname}
\def\subsectionstyle#1{\global\expandafter\let\expandafter\@subsectionstyle
     \csname#1text\endcsname}

\def\chapter#1{%
     \ifdim\lastskip=17sp \else\chapterbreak\vskip\abovechapterskip\fi
     \@chapterstyle{\ifblank\chapternumstyle\then
          \else\newchapternum=\next\chapternumformat\ \fi#1}%
     \nobreak\vskip\belowchapterskip\vskip17sp }

\def\section#1{%
     \ifdim\lastskip=17sp \else\sectionbreak\vskip\abovesectionskip\fi
     \@sectionstyle{\ifblank\sectionnumstyle\then
          \else\newsectionnum=\next\sectionnumformat\ \fi#1}%
     \nobreak\vskip\belowsectionskip\vskip17sp }

\def\subsection#1{%
     \ifdim\lastskip=17sp \else\subsectionbreak\vskip\abovesubsectionskip\fi
     \@subsectionstyle{\ifblank\subsectionnumstyle\then
          \else\newsubsectionnum=\next\subsectionnumformat\ \fi#1}%
     \nobreak\vskip\belowsubsectionskip\vskip17sp }


\let\TeXunderline=\underline
\let\TeXoverline=\overline
\def\underline#1{\relax\ifmmode\TeXunderline{#1}\else
     $\TeXunderline{\hbox{#1}}$\fi}
\def\overline#1{\relax\ifmmode\TeXoverline{#1}\else
     $\TeXoverline{\hbox{#1}}$\fi}

\def\baselinestretch{\afterassignment\@baselinestretch\count@}
\def\@baselinestretch{\baselineskip=\normalbaselineskip
     \divide\baselineskip by\@m\baselineskip=\count@\baselineskip
     \setbox\strutbox=\hbox{\vrule
          height.708\baselineskip depth.292\baselineskip width\z@}%
     \bigskipamount=\the\baselineskip
          plus.25\baselineskip minus.25\baselineskip
     \medskipamount=.5\baselineskip
          plus.125\baselineskip minus.125\baselineskip
     \smallskipamount=.25\baselineskip
          plus.0625\baselineskip minus.0625\baselineskip}

\def\\{\ifhmode\ifnum\lastpenalty=-\@M\else\hfil\penalty-\@M\fi\fi
     \ignorespaces}
\def\newpage{\vfil\break}

\def\lefttext#1{\par{\@text\leftskip=\z@\rightskip=\centering
     \noindent#1\par}}
\def\righttext#1{\par{\@text\leftskip=\centering\rightskip=\z@
     \noindent#1\par}}
\def\centertext#1{\par{\@text\leftskip=\centering\rightskip=\centering
     \noindent#1\par}}
\def\@text{\parindent=\z@ \parfillskip=\z@ \everypar={}%
     \spaceskip=.3333em \xspaceskip=.5em
     \def\\{\ifhmode\ifnum\lastpenalty=-\@M\else\penalty-\@M\fi\fi
          \ignorespaces}}

\def\beginleft{\par\@text\leftskip=\z@ \rightskip=\centering}
     
\def\beginright{\par\@text\leftskip=\centering\rightskip=\z@ }
     
\def\begincenter{\par\@text\leftskip=\centering\rightskip=\centering}

\def\beginnarrow{\defaultoption[\parindent]\@beginnarrow}
\def\@beginnarrow[#1]{\par\advance\leftskip by#1\advance\rightskip by#1}

\begingroup
\catcode`\[=1 \catcode`\{=11 \gdef\beginignore[\endgroup\bgroup
     \catcode`\e=0 \catcode`\\=12 \catcode`\{=11 \catcode`\f=12 \let\or=\relax
     \let\nd{ignor=\fi \let\}=\egroup
     \iffalse}
\endgroup

\long\def\marginnote#1{\leavevmode
     \edef\@marginsf{\spacefactor=\the\spacefactor\relax}%
     \ifdraft\strut\vadjust{%
          \hbox to\z@{\hskip\hsize\hskip.1in
               \vbox to\z@{\vskip-\dp\strutbox
                    \marginnoteformat
                    \vskip-\ht\strutbox
                    \noindent\strut#1\par
                    \vss}%
               \hss}}%
     \fi
     \@marginsf}


\newtoks\everybye \everybye={\par\vfil}
\outer\def\bye{\the\everybye
     \footnotecheck
     \prelabelcheck
     \streamcheck
     \supereject
     \TeXend}

\message{footnotes,}

\newcount\footnotenum \footnotenum=0
\newskip\footnoteskip
\let\@footnotelist=\empty

\def\footnotenumstyle#1{\@setnumstyle\footnotenum{#1}%
     \useafter\ifx{@footnotenumstyle}\symbols
          \global\let\@footup=\empty
     \else\global\let\@footup=\markup
     \fi}

\def\footnote{\footnotecheck\defaultoption[]\@footnote}
\def\@footnote[#1]{\@footnotemark[#1]\@footnotetext}

\def\footnotemark{\defaultoption[]\@footnotemark}
\def\@footnotemark[#1]{\let\@footsf=\empty
     \ifhmode\edef\@footsf{\spacefactor=\the\spacefactor\relax}\/\fi
     \ifnoarg#1\then
          \global\advance\footnotenum by\@ne
          \@footup{\footnotenumformat}%
          \edef\@@foota{\footnotenum=\the\footnotenum\relax}%
          \expandafter\additemR\expandafter\@footup\expandafter
               {\@@foota\footnotenumformat}\to\@footnotelist
          \global\let\@footnotelist=\@footnotelist
     \else\markup{#1}%
          \additemR\markup{#1}\to\@footnotelist
          \global\let\@footnotelist=\@footnotelist
     \fi
     \@footsf}

\def\footnotetext{%
     \ifx\@footnotelist\empty\err@extrafootnotetext\else\@footnotetext\fi}
\def\@footnotetext{%
     \getitemL\@footnotelist\to\@@foota
     \global\let\@footnotelist=\@footnotelist
     \insert\footins\bgroup
     \footnoteformat
     \splittopskip=\ht\strutbox\splitmaxdepth=\dp\strutbox
     \interlinepenalty=\interfootnotelinepenalty\floatingpenalty=\@MM
     \noindent\llap{\@@foota}\strut
     \bgroup\aftergroup\@footnoteend
     \let\@@scratcha=}
\def\@footnoteend{\strut\par\vskip\footnoteskip\egroup}

\def\footnoterule{\normalfonts
     \kern-.3em \hrule width2in height.04em \kern .26em }

\def\footnotecheck{%
     \ifx\@footnotelist\empty
     \else\err@extrafootnotemark
          \global\let\@footnotelist=\empty
     \fi}

\message{labels,}

\let\@@labeldef=\xdef
\newif\if@labelfile
\newwrite\@labelfile
\let\@prelabellist=\empty

\def\label#1#2{\trim#1\to\@@labarg\edef\@@labtext{#2}%
     \edef\@@labname{lab@\@@labarg}%
     \useafter\ifundefined\@@labname\then\else\@yeslab\fi
     \useafter\@@labeldef\@@labname{#2}%
     \ifstreaming
          \expandafter\toks@\expandafter\expandafter\expandafter
               {\csname\@@labname\endcsname}%
          \immediate\write\streamout{\noexpand\label{\@@labarg}{\the\toks@}}%
     \fi}
\def\@yeslab{%
     \useafter\ifundefined{if\@@labname}\then
          \err@labelredef\@@labarg
     \else\useif{if\@@labname}\then
               \err@labelredef\@@labarg
          \else\global\usename{\@@labname true}%
               \useafter\ifundefined{pre\@@labname}\then
               \else\useafter\ifx{pre\@@labname}\@@labtext
                    \else\err@badlabelmatch\@@labarg
                    \fi
               \fi
               \if@labelfile
               \else\global\@labelfiletrue
                    \immediate\write\sixt@@n{--> Creating file \jobname.lab}%
                    \immediate\openout\@labelfile=\jobname.lab
               \fi
               \immediate\write\@labelfile
                    {\noexpand\prelabel{\@@labarg}{\@@labtext}}%
          \fi
     \fi}

\def\putlab#1{\trim#1\to\@@labarg\edef\@@labname{lab@\@@labarg}%
     \useafter\ifundefined\@@labname\then\@nolab\else\usename\@@labname\fi}
\def\@nolab{%
     \useafter\ifundefined{pre\@@labname}\then
          \undefinedlabelformat
          \err@needlabel\@@labarg
          \useafter\xdef\@@labname{\undefinedlabelformat}%
     \else\usename{pre\@@labname}%
          \useafter\xdef\@@labname{\usename{pre\@@labname}}%
     \fi
     \useafter\newif{if\@@labname}%
     \expandafter\additemR\@@labarg\to\@prelabellist}

\def\prelabel#1{\useafter\gdef{prelab@#1}}

\def\ifundefinedlabel#1\then{%
     \expandafter\ifx\csname lab@#1\endcsname\relax}
\def\useiflab#1\then{\csname iflab@#1\endcsname}

\def\prelabelcheck{{%
     \def\^^\##1{\useiflab{##1}\then\else\err@undefinedlabel{##1}\fi}%
     \@prelabellist}}

\message{equation numbering,}

\newcount\chapternum
\newcount\sectionnum
\newcount\subsectionnum
\newcount\equationnum
\newcount\subequationnum
\newcount\figurenum
\newcount\subfigurenum
\newcount\tablenum
\newcount\subtablenum

\newif\if@subeqncount
\newif\if@subfigcount
\newif\if@subtblcount

\def\newchapternum{\newsectionnum=\z@\@resetnum\chapternum}
\def\newsectionnum{\newsubsectionnum=\z@\@resetnum\sectionnum}
\def\newsubsectionnum{\newequationnum=\z@\newfigurenum=\z@\newtablenum=\z@
     \@resetnum\subsectionnum}
\def\newequationnum{\newsubequationnum=\z@\@resetnum\equationnum}
\def\newsubequationnum{\@resetnum\subequationnum}
\def\newfigurenum{\newsubfigurenum=\z@\@resetnum\figurenum}
\def\newsubfigurenum{\@resetnum\subfigurenum}
\def\newtablenum{\newsubtablenum=\z@\@resetnum\tablenum}
\def\newsubtablenum{\@resetnum\subtablenum}

\def\@resetnum#1{\global\advance#1by1 \edef\next{\the#1\relax}\global#1}

\newchapternum=0

\def\chapternumstyle#1{\@setnumstyle\chapternum{#1}}
\def\sectionnumstyle#1{\@setnumstyle\sectionnum{#1}}
\def\subsectionnumstyle#1{\@setnumstyle\subsectionnum{#1}}
\def\equationnumstyle#1{\@setnumstyle\equationnum{#1}}
\def\subequationnumstyle#1{\@setnumstyle\subequationnum{#1}%
     \ifblank\subequationnumstyle\then\global\@subeqncountfalse\fi
     \ignorespaces}
\def\figurenumstyle#1{\@setnumstyle\figurenum{#1}}
\def\subfigurenumstyle#1{\@setnumstyle\subfigurenum{#1}%
     \ifblank\subfigurenumstyle\then\global\@subfigcountfalse\fi
     \ignorespaces}
\def\tablenumstyle#1{\@setnumstyle\tablenum{#1}}
\def\subtablenumstyle#1{\@setnumstyle\subtablenum{#1}%
     \ifblank\subtablenumstyle\then\global\@subtblcountfalse\fi
     \ignorespaces}

\def\eqnlabel#1{%
     \if@subeqncount
          \newsubequationnum=\next
     \else\newequationnum=\next
          \ifblank\subequationnumstyle\then
          \else\global\@subeqncounttrue
               \newsubequationnum=\@ne
          \fi
     \fi
     \label{#1}{\puteqnformat}(\puteqn{#1})%
     \ifdraft\rlap{\hskip.1in{\tt#1}}\fi}

\let\puteqn=\putlab

\def\equation#1#2{\useafter\gdef{eqn@#1}{#2\eqno\eqnlabel{#1}}}
\def\Equation#1{\useafter\gdef{eqn@#1}}

\def\putequation#1{\useafter\ifundefined{eqn@#1}\then
     \err@undefinedeqn{#1}\else\usename{eqn@#1}\fi}

\def\eqnseriesstyle#1{\gdef\@eqnseriesstyle{#1}}
\def\begineqnseries{\subequationnumstyle{\@eqnseriesstyle}%
     \defaultoption[]\@begineqnseries}
\def\@begineqnseries[#1]{\edef\@@eqnname{#1}}
\def\endeqnseries{\subequationnumstyle{blank}%
     \expandafter\ifnoarg\@@eqnname\then
     \else\label\@@eqnname{\puteqnformat}%
     \fi
     \aftergroup\ignorespaces}

\def\figlabel#1{%
     \if@subfigcount
          \newsubfigurenum=\next
     \else\newfigurenum=\next
          \ifblank\subfigurenumstyle\then
          \else\global\@subfigcounttrue
               \newsubfigurenum=\@ne
          \fi
     \fi
     \label{#1}{\putfigformat}\putfig{#1}%
     {\def\marginnoteformat{\tt}\marginnote{#1}}}

\let\putfig=\putlab

\def\figseriesstyle#1{\gdef\@figseriesstyle{#1}}
\def\beginfigseries{\subfigurenumstyle{\@figseriesstyle}%
     \defaultoption[]\@beginfigseries}
\def\@beginfigseries[#1]{\edef\@@figname{#1}}
\def\endfigseries{\subfigurenumstyle{blank}%
     \expandafter\ifnoarg\@@figname\then
     \else\label\@@figname{\putfigformat}%
     \fi
     \aftergroup\ignorespaces}

\def\tbllabel#1{%
     \if@subtblcount
          \newsubtablenum=\next
     \else\newtablenum=\next
          \ifblank\subtablenumstyle\then
          \else\global\@subtblcounttrue
               \newsubtablenum=\@ne
          \fi
     \fi
     \label{#1}{\puttblformat}\puttbl{#1}%
     {\def\marginnoteformat{\tt}\marginnote{#1}}}

\let\puttbl=\putlab

\def\tblseriesstyle#1{\gdef\@tblseriesstyle{#1}}
\def\begintblseries{\subtablenumstyle{\@tblseriesstyle}%
     \defaultoption[]\@begintblseries}
\def\@begintblseries[#1]{\edef\@@tblname{#1}}
\def\endtblseries{\subtablenumstyle{blank}%
     \expandafter\ifnoarg\@@tblname\then
     \else\label\@@tblname{\puttblformat}%
     \fi
     \aftergroup\ignorespaces}

\message{reference numbering,}

\newcount\referencenum \referencenum=0
\newcount\@@prerefcount \@@prerefcount=0
\newcount\@@thisref
\newcount\@@lastref
\newcount\@@loopref
\newcount\@@refseq
\newdimen\refnumindent
\let\@undefreflist=\empty

\def\referencenumstyle#1{\@setnumstyle\referencenum{#1}}

\def\referencestyle#1{\usename{@ref#1}}

\def\@refsequential{%
     \gdef\@refpredef##1{\global\advance\referencenum by\@ne
          \let\^^\=0\label{##1}{\^^\{\the\referencenum}}%
          \useafter\gdef{ref@\the\referencenum}{{##1}{\undefinedlabelformat}}}%
     \gdef\@reference##1##2{%
          \ifundefinedlabel##1\then
          \else\def\^^\####1{\global\@@thisref=####1\relax}\putlab{##1}%
               \useafter\gdef{ref@\the\@@thisref}{{##1}{##2}}%
          \fi}%
     \gdef\endputreferences{%
          \loop\ifnum\@@loopref<\referencenum
                    \advance\@@loopref by\@ne
                    \expandafter\expandafter\expandafter\@printreference
                         \csname ref@\the\@@loopref\endcsname
          \repeat
          \par}}

\def\@refpreordered{%
     \gdef\@refpredef##1{\global\advance\referencenum by\@ne
          \additemR##1\to\@undefreflist}%
     \gdef\@reference##1##2{%
          \ifundefinedlabel##1\then
          \else\global\advance\@@loopref by\@ne
               {\let\^^\=0\label{##1}{\^^\{\the\@@loopref}}}%
               \@printreference{##1}{##2}%
          \fi}
     \gdef\endputreferences{%
          \def\^^\####1{\useiflab{####1}\then
               \else\reference{####1}{\undefinedlabelformat}\fi}%
          \@undefreflist
          \par}}

\def\beginprereferences{\par
     \def\reference##1##2{\global\advance\referencenum by1\@ne
          \let\^^\=0\label{##1}{\^^\{\the\referencenum}}%
          \useafter\gdef{ref@\the\referencenum}{{##1}{##2}}}}
\def\endprereferences{\global\@@prerefcount=\the\referencenum\par}

\def\beginputreferences{\par
     \refnumindent=\z@\@@loopref=\z@
     \loop\ifnum\@@loopref<\referencenum
               \advance\@@loopref by\@ne
               \setbox\z@=\hbox{\referencenum=\@@loopref
                    \referencenumformat\enskip}%
               \ifdim\wd\z@>\refnumindent\refnumindent=\wd\z@\fi
     \repeat
     \putreferenceformat
     \@@loopref=\z@
     \loop\ifnum\@@loopref<\@@prerefcount
               \advance\@@loopref by\@ne
               \expandafter\expandafter\expandafter\@printreference
                    \csname ref@\the\@@loopref\endcsname
     \repeat
     \let\reference=\@reference}

\def\@printreference#1#2{\ifx#2\undefinedlabelformat\err@undefinedref{#1}\fi
     \noindent\ifdraft\rlap{\hskip\hsize\hskip.1in \tt#1}\fi
     \llap{\referencenum=\@@loopref\referencenumformat\enskip}#2\par}

\def\reference#1#2{{\par\refnumindent=\z@\putreferenceformat\noindent#2\par}}

\def\putref#1{\trim#1\to\@@refarg
     \expandafter\ifnoarg\@@refarg\then
          \toks@={\relax}%
     \else\@@lastref=-\@m\def\@@refsep{}\def\@more{\@nextref}%
          \toks@={\@nextref#1,,}%
     \fi\the\toks@}
\def\@nextref#1,{\trim#1\to\@@refarg
     \expandafter\ifnoarg\@@refarg\then
          \let\@more=\relax
     \else\ifundefinedlabel\@@refarg\then
               \expandafter\@refpredef\expandafter{\@@refarg}%
          \fi
          \def\^^\##1{\global\@@thisref=##1\relax}%
          \global\@@thisref=\m@ne
          \setbox\z@=\hbox{\putlab\@@refarg}%
     \fi
     \advance\@@lastref by\@ne
     \ifnum\@@lastref=\@@thisref\advance\@@refseq by\@ne\else\@@refseq=\@ne\fi
     \ifnum\@@lastref<\z@
     \else\ifnum\@@refseq<\thr@@
               \@@refsep\def\@@refsep{,}%
               \ifnum\@@lastref>\z@
                    \advance\@@lastref by\m@ne
                    {\referencenum=\@@lastref\putrefformat}%
               \else\undefinedlabelformat
               \fi
          \else\def\@@refsep{--}%
          \fi
     \fi
     \@@lastref=\@@thisref
     \@more}

\message{streaming,}

\newif\ifstreaming

\def\streamto{\defaultoption[\jobname]\@streamto}
\def\@streamto[#1]{\global\streamingtrue
     \immediate\write\sixt@@n{--> Streaming to #1.str}%
     \newwrite\streamout\immediate\openout\streamout=#1.str }

\def\streamfrom{\defaultoption[\jobname]\@streamfrom}
\def\@streamfrom[#1]{\newread\streamin\openin\streamin=#1.str
     \ifeof\streamin
          \expandafter\err@nostream\expandafter{#1.str}%
     \else\immediate\write\sixt@@n{--> Streaming from #1.str}%
          \let\@@labeldef=\gdef
          \ifstreaming
               \edef\@elc{\endlinechar=\the\endlinechar}%
               \endlinechar=\m@ne
               \loop\read\streamin to\@@scratcha
                    \ifeof\streamin
                         \streamingfalse
                    \else\toks@=\expandafter{\@@scratcha}%
                         \immediate\write\streamout{\the\toks@}%
                    \fi
                    \ifstreaming
               \repeat
               \@elc
               \input #1.str
               \streamingtrue
          \else\input #1.str
          \fi
          \let\@@labeldef=\xdef
     \fi}

\def\streamcheck{\ifstreaming
     \immediate\write\streamout{\pagenum=\the\pagenum}%
     \immediate\write\streamout{\footnotenum=\the\footnotenum}%
     \immediate\write\streamout{\referencenum=\the\referencenum}%
     \immediate\write\streamout{\chapternum=\the\chapternum}%
     \immediate\write\streamout{\sectionnum=\the\sectionnum}%
     \immediate\write\streamout{\subsectionnum=\the\subsectionnum}%
     \immediate\write\streamout{\equationnum=\the\equationnum}%
     \immediate\write\streamout{\subequationnum=\the\subequationnum}%
     \immediate\write\streamout{\figurenum=\the\figurenum}%
     \immediate\write\streamout{\subfigurenum=\the\subfigurenum}%
     \immediate\write\streamout{\tablenum=\the\tablenum}%
     \immediate\write\streamout{\subtablenum=\the\subtablenum}%
     \immediate\closeout\streamout
     \fi}


\def\err@badtypesize{%
     \errhelp={The limited availability of certain fonts requires^^J%
          that the base type size be 10pt, 12pt, or 14pt.^^J}%
     \errmessage{--> Illegal base type size}}

\def\err@badsizechange{\immediate\write\sixt@@n
     {--> Size change not allowed in math mode, ignored}}

\def\err@sizetoolarge#1{\immediate\write\sixt@@n
     {--> \noexpand#1 too big, substituting HUGE}}

\def\err@sizenotavailable#1{\immediate\write\sixt@@n
     {--> Size not available, \noexpand#1 ignored}}

\def\err@fontnotavailable#1{\immediate\write\sixt@@n
     {--> Font not available, \noexpand#1 ignored}}

\def\err@sltoit{\immediate\write\sixt@@n
     {--> Style \noexpand\sl not available, substituting \noexpand\it}%
     \it}

\def\err@bfstobf{\immediate\write\sixt@@n
     {--> Style \noexpand\bfs not available, substituting \noexpand\bf}%
     \bf}

\def\err@badgroup#1#2{%
     \errhelp={The block you have just tried to close was not the one^^J%
          most recently opened.^^J}%
     \errmessage{--> \noexpand\end{#1} doesn't match \noexpand\begin{#2}}}

\def\err@badcountervalue#1{\immediate\write\sixt@@n
     {--> Counter (#1) out of bounds}}

\def\err@extrafootnotemark{\immediate\write\sixt@@n
     {--> \noexpand\footnotemark command
          has no corresponding \noexpand\footnotetext}}

\def\err@extrafootnotetext{%
     \errhelp{You have given a \noexpand\footnotetext command without first
          specifying^^Ja \noexpand\footnotemark.^^J}%
     \errmessage{--> \noexpand\footnotetext command has no corresponding
          \noexpand\footnotemark}}

\def\err@labelredef#1{\immediate\write\sixt@@n
     {--> Label "#1" redefined}}

\def\err@badlabelmatch#1{\immediate\write\sixt@@n
     {--> Definition of label "#1" doesn't match value in \jobname.lab}}

\def\err@needlabel#1{\immediate\write\sixt@@n
     {--> Label "#1" cited before its definition}}

\def\err@undefinedlabel#1{\immediate\write\sixt@@n
     {--> Label "#1" cited but never defined}}

\def\err@undefinedeqn#1{\immediate\write\sixt@@n
     {--> Equation "#1" not defined}}

\def\err@undefinedref#1{\immediate\write\sixt@@n
     {--> Reference "#1" not defined}}

\def\err@nostream#1{%
     \errhelp={You have tried to input a stream file that doesn't exist.^^J}%
     \errmessage{--> Stream file #1 not found}}

\message{jyTeX initialization}

\everyjob{\immediate\write16{--> jyTeX version \fmtversion}%
     \edef\@@jobname{\jobname}%
     \edef\jobname{\@@jobname}%
     \settime
     \openin0=\jobname.lab
     \ifeof0
     \else\closein0
          \immediate\write16{--> Getting labels from file \jobname.lab}%
          \input\jobname.lab
     \fi}


\def\fixedskipslist{%
     \^^\{\topskip}%
     \^^\{\splittopskip}%
     \^^\{\maxdepth}%
     \^^\{\skip\topins}%
     \^^\{\skip\footins}%
     \^^\{\headskip}%
     \^^\{\footskip}}

\def\scalingskipslist{%
     \^^\{\p@renwd}%
     \^^\{\delimitershortfall}%
     \^^\{\nulldelimiterspace}%
     \^^\{\scriptspace}%
     \^^\{\jot}%
     \^^\{\normalbaselineskip}%
     \^^\{\normallineskip}%
     \^^\{\normallineskiplimit}%
     \^^\{\baselineskip}%
     \^^\{\lineskip}%
     \^^\{\lineskiplimit}%
     \^^\{\bigskipamount}%
     \^^\{\medskipamount}%
     \^^\{\smallskipamount}%
     \^^\{\parskip}%
     \^^\{\parindent}%
     \^^\{\abovedisplayskip}%
     \^^\{\belowdisplayskip}%
     \^^\{\abovedisplayshortskip}%
     \^^\{\belowdisplayshortskip}%
     \^^\{\abovechapterskip}%
     \^^\{\belowchapterskip}%
     \^^\{\abovesectionskip}%
     \^^\{\belowsectionskip}%
     \^^\{\abovesubsectionskip}%
     \^^\{\belowsubsectionskip}}


\def\twoupsetup{
     \topmargin=.75in
     \leftmargin=.5in
     \vsize=6.9in
     \hsize=4.75in
     \fullhsize=10in
     \let\draft=\relax}

\outputstyle{normal}                             

\def\marginnoteformat{\subscriptsize             
     \hsize=1in \baselinestretch=1000 \everypar={}%
     \tolerance=5000 \hbadness=5000 \parskip=0pt \parindent=0pt
     \leftskip=0pt \rightskip=0pt \raggedright}

\head={\ifdraft\normalfonts\it\hfil DRAFT\hfil   
     \llap{\number\day\ \monthword\month\ \militarytime}\else\hfil\fi}
\foot={\hfil\normalfonts\numstyle\pagenum\hfil}  

\normalbaselineskip=12pt                         
\normallineskip=0pt                              
\normallineskiplimit=0pt                         
\normalbaselines                                 

\topskip=.85\baselineskip \splittopskip=\topskip \headskip=2\baselineskip
\footskip=\headskip

\pagenumstyle{arabic}                            

\parskip=0pt                                     
\parindent=20pt                                  

\baselinestretch=1000                            


\chapterstyle{left}                              
\chapternumstyle{blank}                          
\def\chapterbreak{\newpage}                      
\abovechapterskip=0pt                            
\belowchapterskip=1.5\baselineskip               
     plus.38\baselineskip minus.38\baselineskip
\def\chapternumformat{\numstyle\chapternum.}     

\sectionstyle{left}                              
\sectionnumstyle{blank}                          
\def\sectionbreak{\vskip0pt plus4\baselineskip\penalty-100
     \vskip0pt plus-4\baselineskip}              
\abovesectionskip=1.5\baselineskip               
     plus.38\baselineskip minus.38\baselineskip
\belowsectionskip=\the\baselineskip              
     plus.25\baselineskip minus.25\baselineskip
\def\sectionnumformat{
     \ifblank\chapternumstyle\then\else\numstyle\chapternum.\fi
     \numstyle\sectionnum.}

\subsectionstyle{left}                           
\subsectionnumstyle{blank}                       
\def\subsectionbreak{\vskip0pt plus4\baselineskip\penalty-100
     \vskip0pt plus-4\baselineskip}              
\abovesubsectionskip=\the\baselineskip           
     plus.25\baselineskip minus.25\baselineskip
\belowsubsectionskip=.75\baselineskip            
     plus.19\baselineskip minus.19\baselineskip
\def\subsectionnumformat{
     \ifblank\chapternumstyle\then\else\numstyle\chapternum.\fi
     \ifblank\sectionnumstyle\then\else\numstyle\sectionnum.\fi
     \numstyle\subsectionnum.}


\footnotenumstyle{symbols}                       
\footnoteskip=0pt                                
\def\footnotenumformat{\numstyle\footnotenum}    
\def\footnoteformat{\footnotesize                
     \everypar={}\parskip=0pt \parfillskip=0pt plus1fil
     \leftskip=1em \rightskip=0pt
     \spaceskip=0pt \xspaceskip=0pt
     \def\\{\ifhmode\ifnum\lastpenalty=-10000
          \else\hfil\penalty-10000 \fi\fi\ignorespaces}}


\def\undefinedlabelformat{$\bullet$}             


\equationnumstyle{arabic}                        
\subequationnumstyle{blank}                      
\figurenumstyle{arabic}                          
\subfigurenumstyle{blank}                        
\tablenumstyle{arabic}                           
\subtablenumstyle{blank}                         

\eqnseriesstyle{alphabetic}                      
\figseriesstyle{alphabetic}                      
\tblseriesstyle{alphabetic}                      

\def\puteqnformat{\hbox{
     \ifblank\chapternumstyle\then\else\numstyle\chapternum.\fi
     \ifblank\sectionnumstyle\then\else\numstyle\sectionnum.\fi
     \ifblank\subsectionnumstyle\then\else\numstyle\subsectionnum.\fi
     \numstyle\equationnum
     \numstyle\subequationnum}}
\def\putfigformat{\hbox{
     \ifblank\chapternumstyle\then\else\numstyle\chapternum.\fi
     \ifblank\sectionnumstyle\then\else\numstyle\sectionnum.\fi
     \ifblank\subsectionnumstyle\then\else\numstyle\subsectionnum.\fi
     \numstyle\figurenum
     \numstyle\subfigurenum}}
\def\puttblformat{\hbox{
     \ifblank\chapternumstyle\then\else\numstyle\chapternum.\fi
     \ifblank\sectionnumstyle\then\else\numstyle\sectionnum.\fi
     \ifblank\subsectionnumstyle\then\else\numstyle\subsectionnum.\fi
     \numstyle\tablenum
     \numstyle\subtablenum}}


\referencestyle{sequential}                      
\referencenumstyle{arabic}                       
\def\putrefformat{\numstyle\referencenum}        
\def\referencenumformat{\numstyle\referencenum.} 
\def\putreferenceformat{
     \everypar={\hangindent=1em \hangafter=1 }%
     \def\\{\hfil\break\null\hskip-1em \ignorespaces}%
     \leftskip=\refnumindent\parindent=0pt \interlinepenalty=1000 }


\normalsize


\def\fmtversion{2.6M (June 1992)}

\catcode`\@=12

\typesize=10pt \magnification=1200 \baselineskip17truept
\footnotenumstyle{arabic} \hsize=6truein\vsize=8.5truein
\input epsf
\sectionnumstyle{blank}
\chapternumstyle{blank}
\chapternum=1
\sectionnum=1
\pagenum=0

\def\begintitle{\pagenumstyle{blank}\parindent=0pt
\begin{narrow}[0.4in]}
\def\endtitle{\end{narrow}\newpage\pagenumstyle{arabic}}


\def\beginexercise{\vskip 20truept\parindent=0pt\begin{narrow}[10
truept]}
\def\endexercise{\vskip 10truept\end{narrow}}


\def\eql#1{\eqno\eqnlabel{#1}}
\def\ref{\reference}
\def\peq{\puteqn}
\def\pref{\putref}

\def\mgn{\marginnote}
\def\bex{\begin{exercise}}
\def\eex{\end{exercise}}


\font\open=msbm10 


\def\StretchRtArr#1{{\count255=0\loop\relbar\joinrel\advance\count255 by1
\ifnum\count255<#1\repeat\rightarrow}}
\def\StretchLtArr#1{\,{\leftarrow\!\!\count255=0\loop\relbar
\joinrel\advance\count255 by1\ifnum\count255<#1\repeat}}

\def\StretchLRtArr#1{\,{\leftarrow\!\!\count255=0\loop\relbar\joinrel\advance
\count255 by1\ifnum\count255<#1\repeat\rightarrow\,\,}}

\def\mbox#1{{\leavevmode\hbox{#1}}}

\def\hspace#1{{\phantom{\mbox#1}}}
\def\oZ{\mbox{\open\char90}}

\def\oN{\mbox{\open\char78}}

\def\al{\alpha}
\def\bom{{\bmit\omega}}
\def\be{\beta}

\def\de{\delta}
\def\Ga{\Gamma}

\def\ep{\epsilon}

\def\la{\lambda}
\def\La{\Lambda}
\def\om{\omega}
\def\Om{\Omega}

\def\si{\sigma}
\def\Si{\Sigma}
\def\th{\theta}

\def\ze{\zeta}

\def\De{\Delta}

\def\caA{{\cal A}}
\def\caC{{\cal C}}

\def\caG{{\cal G}}

\def\caO{{\cal O}}

\def\Real{{\rm Re\,}}

\def\sc{{\rm sc }}

\def\zf{$\zeta$--function}
\def\zfs{$\zeta$--functions}


\def\frac#1/#2{\leavevmode\kern.1em
\raise.5ex\hbox{\the\scriptfont0 #1}\kern-.1em/\kern-.15em
\lower.25ex\hbox{\the\scriptfont0 #2}}
\def\sfrac#1/#2{\leavevmode\kern.1em
\raise.5ex\hbox{\the\scriptscriptfont0 #1}\kern-.1em/\kern-.15em
\lower.25ex\hbox{\the\scriptscriptfont0 #2}}

\def\gtorder{\mathrel{\raise.3ex\hbox{$>$}\mkern-14mu
             \lower0.6ex\hbox{$\sim$}}}
\def\ltorder{\mathrel{\raise.3ex\hbox{$<$}\mkern-14mu
             \lower0.6ex\hbox{$\sim$}}}

\def\semidirprod{\rlap{\ss C}\raise1pt\hbox{$\mkern.75mu\times$}}
\def\for{\lower6pt\hbox{$\Big|$}}
\def\fish{\kern-.25em{\phantom{abcde}\over \phantom{abcde}}\kern-.25em}


\def\boxit#1{\vbox{\hrule\hbox{\vrule\kern3pt
        \vbox{\kern3pt#1\kern3pt}\kern3pt\vrule}\hrule}}
\def\dalemb#1#2{{\vbox{\hrule height .#2pt
        \hbox{\vrule width.#2pt height#1pt \kern#1pt \vrule
                width.#2pt} \hrule height.#2pt}}}

\def\ol{\overline}
\def\frac#1#2{{{#1}\over{#2}}}

\def\noin{\noindent}

\def\comb#1#2{{\left(#1\atop#2\right)}}

\def\cosech{{\rm cosech\,}}

\def\sech{{\rm sech\,}}

\def\eg{{\it e.g.}}
\def\ie{{\it i.e. }}
\def\cf{{\it cf }}
\def\pa{\partial}

\def\av#1{\langle#1\rangle} 


\def\3j#1#2#3#4#5#6{\left\lgroup\matrix{#1&#2&#3\cr#4&#5&#6\cr}
\right\rgroup}

\def\m?{\mgn{?}}

\def\pa{\partial}

\def\beq{\begin{eqnarray}}
\def\eeq{\end{eqnarray}}


\def\aop#1#2#3{{\it Ann. Phys.} {\bf {#1}} ({#2}) #3}

\def\cmp#1#2#3{{\it Comm. Math. Phys.} {\bf {#1}} ({#2}) #3}

\def\jmp#1#2#3{{\it J. Math. Phys.} {\bf {#1}} ({#2}) #3}
\def\jpa#1#2#3{{\it J. Phys.} {\bf A{#1}} ({#2}) #3}

\def\np#1#2#3{{\it Nucl. Phys.} {\bf B{#1}} ({#2}) #3}

\def\prB#1#2#3{{\it Phys. Rev.} {\bf B{#1}} ({#2}) #3}
\def\prD#1#2#3{{\it Phys. Rev.} {\bf D{#1}} ({#2}) #3}
\def\prl#1#2#3{{\it Phys. Rev. Lett.} {\bf #1} ({#2}) #3}

\def\am#1#2#3{{\it Acta Mathematica} {\bf {#1}} ({#2}) #3}

\def\jpamt#1#2#3{{\it J. Phys.A:Math.Theor.} {\bf{#1}} ({#2}) #3}
\def\jram#1#2#3{{\it J. f. reine u. Angew. Math.} {\bf {#1}} ({#2}) #3}

\def\mz#1#2#3{{\it Math. Zeit.} {\bf {#1}} ({#2}) #3}

\def\plb#1#2#3{{\it Phys. Letts.} {\bf {B#1}} ({#2}) #3}

\def\qjm#1#2#3{{\it Quart. J. Math.} {\bf {#1}} ({#2}) #3}

\input renyichempot1.lab
\begin{title}
\vglue 0.5truein
\vskip15truept
\centertext {\Bigfonts \bf Charged R\'enyi entropies} \vskip7truept
\vskip10truept\centertext{\Bigfonts \bf for free scalar fields} \vskip17truept
\centertext{\Bigfonts \bf }
 \vskip 20truept
\centertext{J.S.Dowker\footnote{ dowker@man.ac.uk;  dowkeruk@yahoo.co.uk}} \vskip
7truept \centertext{\it Theory Group,} \centertext{\it School of Physics and Astronomy,}
\centertext{\it The University of Manchester,} \centertext{\it Manchester, England} \vskip
7truept \centertext{}

\vskip 7truept

\vskip40truept
\begin{narrow}
I first calculate the charged spherical R\'enyi entropy by a numerical method that does not
require knowledge of any eigenvalue degeneracies, and applies to all odd dimensions.

An image method is used to relate the full sphere values to those for an integer covering,
$n$. It is shown to be equivalent to a `transformation' property of the zeta--function.

The $n\to\infty$ limit is explicitly constructed analytically and a relation deduced between
the limits of corner coefficients and the effective action (free energy) which generalises,
for free fields, a result of Bueno, Myers and Witczak--Krempa and Elvang and Hadjiantonis
to any dimension.

Finally, the known polynomial expressions for the R\'enyi entropy on even spheres at zero
chemical potential are re--derived in a different form and a simple formula for the
conformal anomaly given purely in terms of central factorials is obtained.

\end{narrow}
\vskip 5truept
\vskip 60truept
\vfil
\end{title}
\pagenum=0
\newpage

\section{\bf 1. Introduction and summary.}
In a recent work I discussed one aspect of charged spherical R\'enyi entropy viz. the
conformal weights of twist operators. I now turn to the R\'enyi entropies themselves, again
for free fields, in this paper just conformal scalars, extending the work in [\pref{B}] to
arbitrary dimensions. Although the object of the many papers around this topic are results
that apply to any conformal field theory, the case of free fields in special geometries is often
used as a valuable exemplar of techniques and putative theorems.

The calculation of the entropy, in this case, can be effected by conformal transformation to
an orbifolded sphere, as used earlier many times. For {\it charged} entropies such an
approach is described in [\pref{B}] using a mode method. My analysis differs somewhat.

The entropy is determined from the effective action, or logdet, of the relevant propagating
operator. Technically one needs the derivative of the corresponding \zf, $\ze(s)$, at $s=0$
which requires the eigenvalues. The uncharged eigenproblem is classic and has been given
many times. In the next section I detail the charged extension before outlining its
application to the entropy, spending most time on the derivation of the effective action
and its numerical evaluation the results of which are shown in graphs.

In a following section, an image relation, derived elsewhere, is given and an explicit
illustrative example calculated. The image expression is then shown, in the spherical
situation, to be a consequence of a `transformation' property of the \zf.

The next section contains a different formulation of the effective action in terms of Barnes'
multiple $\Ga$--function, $\Ga_d$, which, for the full sphere in odd dimensions, reduces to
an integral involving elementary functions and gives a more explicit dependence on the
chemical potential.

A further section has a very brief description of another method involving $\Ga_d$ that
leads to an integral over Gauss' $\psi$ function.

Next, the limit $n\to \infty$ is analysed where $n$ is the covering integer, and, finally, the
R\'enyi entropy on even spheres is reconsidered and the exact expression derived in a
neater form. Also a compact, and easily calculated, formula is obtained for the conformal
anomaly.

\section{\bf 2. The orbifolded sphere with flux}

I deal with the $q$--orbifolded $d$--sphere, S$^d/\oZ_q$ which, in order to allow for a
chemical potential, has an Aharonov--Bohm flux running between the south and north
poles, in either direction equivalently.

The orbifolded sphere is made up of $2q$ segments of apex angle $\pi/q$. I refer to these
as lunes. The combination of two adjacent lunes I call a {\it doubled} lune. These tile the
sphere under an SO(2) action, neighbouring lunes being related by a reflection.

It is convenient to set $\be=\pi/q$ and generalise to a lune of {\it arbitrary} angle. The
tiling property is then geometrically lost.

The $d$--lune can be defined inductively by giving its metric in the nested form,
  $$
  ds^2_{d-lune}=d\th_d^2+\sin^2\th_d\,\,ds^2_{(d-1)-lune}\,,\quad
  0\le \th_i\le\pi\,,\quad i\ne1\,,
  \eql{lune}
  $$
which is iterated down to the $1$--lune of metric $d\th_1^2$ with $0\le\th_1\le\be$. The
angle $\th_1$ is referred to as the polar angle and conventionally written $\phi$. The
angle of the lune (there are two of them) is $\be$. The conical deformation of the sphere
can be traced to the deformation of the $\phi$ circle. This disappears when $\be=2\pi$.

The boundary of the lune comprises two pieces corresponding to $\phi=0$ and $\phi=\be$.
The metric (\peq{lune}) shows immediately that these are unit $(d-1)$--hemispheres
because {\it their} polar angle, $\th_2$, runs only from $0$ to $\pi$. Conditions, typically
Dirichlet and Neumann, can be applied at the boundary. The boundary parts intersect, with
a constant dihedral angle of $\be$, in a $(d-2)$--sphere, of unit radius, which constitutes
a set of points fixed under O(2) rotations parametrised by $\phi$.

It can be seen that the 2--lune submanifold, with coordinates $\th_1$ and $\phi$, has a
wedge singularity at its north and south poles. These poles are at $\th_2=0$ and
$\th_2=\pi$ and are the 0--hemispheres of a 0--sphere. In the $d$--lune, the
submanifolds, $\th_2=0$ and $\th_2=\pi$, are the $(d-2)$--hemispheres of the
$(d-2)$--sphere of fixed points just mentioned and is the entangling surface.

In the absence of the flux, the mode problem is more or less standard \footnote{ For the
full sphere it was first given by Green who, for some reason, works in $d$ dimensions at a
time when the notion of higher dimensional geometry had yet to be formulated.}. I
consider the propagating operator and eigenvalue equation,
  $$\eqalign{
  \caO&=-\De_2+c^2\cr
  \caO\psi&=\La\psi\,,\quad -\De_2\psi=\la\psi\,,\quad \La=\la+c^2\,,
  }
  $$
where $\De_2$ is the Laplacian and $c$ is a constant. The separation of variables follows
the nesting structure in (\peq{lune}) and is a classic calculation. The hyperspherical
harmonics are labeled by the set of $d$ separation constants $l_i$, $i=1,\ldots, d$. The
first one, $l_1$, is associated with the $\phi$ eigenvalue equation, and I leave it free for
the moment. The remaining constants, $l_i$, which can be taken as non--negative, are
associated with the intervening spheres, S$^i$. They satisfy,
  $$
  l_i=l_{i-1}+n_i\,,\quad n_i=0,1,\ldots,\infty\,,\quad i=2,\ldots,d\,,
  \eql{cond2}
  $$
with the last one determining the eigenvalue $\la$ by,
  $$
  \la=l_d\big(l_d+d-1\big)=\big(l_d+{d-1\over2}\big)^2-{(d-1)^2\over4}\,.
  \eql{eigv}
  $$

Iteration of (\peq{cond2}) gives
  $$
  \l_d=l_1+n_2+\ldots+n_d\,,
  \eql{iter}
  $$

In this approach the degeneracy of a particular eigenvalue is given by coincidences as the
free labels vary. If $l_1$ is an integer then the distribution of $d$ integers is involved,
otherwise it is that of $d-1$ integers. It will turn out unnecessary  to find the degeneracies
explicitly, although it is a standard situation.

The calculation of $\la$ devolves upon the constant $l_1$  coming from the eigenproblem
on the deformed $\phi$ circle with a flux running through, an old question \cf
[\pref{DandB}]. I approach this via pseudo--periodic eigenfunctions which satisfy the
defining equation, for a complex $\psi$,
  $$
  {d^2\over d\phi^2}\,\psi_{l_1}(\phi)=l_1^2\,\psi_{l_1}(\phi)\,,
  $$
with the periodicity condition on the {\it double} lune,
  $$
  \psi(\phi+2\be)=e^{2\pi\,i\,\de}\,\psi(\phi)\,.
  \eql{pf}
  $$

Then the eigenfunction is,
  $$
  \psi_{l_1}(\phi)\propto e^{i\,l_1\phi}\,,
  $$
with
  $$
  l_1={\pi\over\be}\big(n_1+\de\big)\,,\quad n_1=-\infty,\ldots 0,\ldots,\infty\,,
  \eql{el}
  $$
and all quantities are periodic in $\de$ with period 1. Furthermore all real physical quantities
are unchanged under the reversal $\de\to-\de$. As remarked in [\pref{Dowcascone}], for
complex fields the Green function, as a typical quantity, is
$G_{2\be,\,\de}+G_{2\be,\,-\de}$ where $G_{2\be,\de}$ is computed from fields obeying
(\peq{pf}). These two conditions require that the periodising unit cell be symmetrical about
zero, $-1/2<\de\le1/2$. My basic approach is to calculate the physical quantities in a unit
cell and simply extend them by periodicity.

Generally, one might expect the physical quantity to have singularities at the cell
boundary, and its iterates. The present calculation shows that this not so, agreeing with
[\pref{B}]. The unit cell could then be chosen $0\le\de<1$.

I turn now to the eigenvalues and note that $l_1$ can be negative but because the
Legendre equation for the next spherical function involves $l_1^2$ it is sufficient to
replace $l_1$ by $|l_1|$ so that the expression for $l_d$, (\peq{iter}), takes the form,
  $$
  l_d=q|n_1\pm \,\de|+{\bf 1}{\bf . n}\,,
  $$
where $q=\pi/\be$ and ${\bf1}$  is a $d-1$--dimensional set with
${\bf1}=\big(1,1,\ldots,1\big)$.

Using (\peq{eigv}), the eigenvalue $\La$ is,
  $$\eqalign{
\La(\pm\de)&=\big({d-1\over2}+q|n_1\pm \,\de|+{\bf1}{\bf . n}\big)^2-
{(d-1)^2\over4}-c^2\,.\cr
}
  $$
For conformal invariance in $d$ dimensions,
  $$\eqalign{
\La(\pm\de)&=\big({d-1\over2}+q|n_1\pm\,\de|+{\bf1}{\bf . n}\big)^2-{1\over4}\,,\cr
}
  $$
and the conformal \zf\ for the operator $\caO$ is then ($\al=1/2$), after some
rearrangement and assuming that $|\de|\le1$,
  $$\eqalign{
   \ze\big(s,\mu,q\mid\bom\big)&=\sum_{\pm}\sum_{n_1=-\infty}^\infty\sum_{\bf n=0}^{\bf\infty}
   \bigg[\big({d-1\over2}+q|n_1\pm\,\de|+{\bf1}{\bf . n}\big)^2-\al^2\bigg]^{-s}\cr
&=2\sum_{\bf n=0}^{\bf\infty}
   \bigg[\big({d-1\over2}+ |\mu|+{\bom}{\bf . n}\big)^2-\al^2\bigg]^{-s}\cr
&+2\sum_{\bf n=0}^{\bf\infty}
   \bigg[\big({d-1\over2}+q-|\mu|+{\bom}{\bf . n}\big)^2-\al^2\bigg]^{-s}\,,\cr
   }
   \eql{zf1}
  $$
where $\bom$ is the set of $d$ numbers, $(q,\bf1)$, but can be considered, formally, as
real non--negative and referred to as the {\it parameters} or {\it degrees}.

I have introduced the chemical potential $\mu=q\de$. In particular, for the orbifolded
sphere, S$^d/\oZ_q$, $q$ is integral, otherwise not. \footnote{ If $q=1/n$ ($n\in\oN$),
$\de$ is the flux through the $n$--fold covering so that $\de/n$ is the flux through a
single sheet \ie the chemical potential as defined here, and in [\pref{B}].} The range of
$\mu$ is extended by periodicity.

When $\mu=0$ the first sum  corresponds to the Neumann lune and the second to the
Dirichlet one.

The factor of 2 in (\peq{zf1}) is a consequence of the charged nature of the field and so
the effective action (sometimes called the free energy), is,
  $$
  \caA_d(q,\mu)=-{1\over2}\ze'\big(0,\mu,q\mid\bom\big)\,.
  $$
Another notation is $\caA_{n,d}(\mu)\equiv \caA_d(1/n,\mu)$.

\begin{ignore}
\bigg[In more detail, essentially, we need, (assume  $-1\le\de\le1$),
  $$\eqalign{
\sum_{\pm}\sum_{n=-\infty}^\infty f(|n\pm\de|&= \sum_{\pm}\bigg(\sum_{n=1}^\infty
f(|n\pm\de|)+\sum_{n=-\infty}^{-1} f(|n\pm\de|) +\,f(|\de|) \bigg)\cr
&=\sum_{\pm}\bigg(\sum_{n=1}^\infty
f(|n\pm\de|)+\sum_{n=1}^{\infty} f(|-n\pm\de|) +\,f(|\de|)\bigg)\cr
&=\sum_{\pm}\bigg(\sum_{n=1}^\infty
f(|n\pm\de|)+\sum_{n=1}^{\infty} f(|n\mp\de|) +\,f(|\de|)\bigg)\cr
&=2\sum_{\pm}\bigg(\sum_{n=1}^\infty
f(|n\pm\de|) +\,f(|\de|)\bigg)\cr
&=2\sum_{\pm}\bigg(\sum_{n=1}^\infty
f(n\pm\de) +\,f(|\de|)\bigg)\cr
}
  $$
using the restriction on $\de$, to give $n\pm\de\ge0$ for $n\ge1$. Now say $\de>0$ then
combining the last term with the $n+\de$ sum,
  $$\eqalign{
\sum_{\pm}\sum_{n=-\infty}^\infty f(|n\pm\de|
&=2\sum_{\pm}\bigg(\sum_{n=1}^\infty
f(n\pm\de) +\,f(\de)\bigg)\cr
&=2\bigg(\sum_{n=0}^\infty
f(n+\de)+\sum_{n=1}^\infty
f(n-\de) \bigg)\cr
}
  $$
and if $\de<0$, likewise
  $$
  =2\bigg(\sum_{n=0}^\infty
f(n+1+\de)+\sum_{n=0}^\infty
f(n-\de) \bigg)
  $$
Hence the answer,
$$\eqalign{
\sum_{\pm}\sum_{n=-\infty}^\infty f(|n\pm\de|
&=2\bigg(\sum_{n=0}^\infty
f(n+|\de|)+\sum_{n=0}^\infty
f(n+1-|\de|) \bigg)\cr
&=2\bigg(\sum_{n=0}^\infty
f(n+1/2+\ol\de)+\sum_{n=0}^\infty
f(n+1/2-\ol\de) \bigg)\cr
}
  $$
where $\ol\de\equiv|\de|-1/2$ and shows symmetry under $\ol\de\to -\ol\de$. Note that
$\de$ can equal $\pm1$.
\end{ignore}

A means of calculating the derivative $\ze'(0)$ is presented in [\pref{Dowcmp}] which I
will employ later but first I give a numerical method used, \eg, in [\pref{Dowren}] and
valid only in odd dimensions.

Since the references and technicalities of the method are detailed in [\pref{Dowren}] I
feel I can just give the answer which is, after minor manipulation,
  $$\eqalign{
  \caA_d(q,\mu)&=-{1\over2^{d-2}}\,\int_0^\infty dx\,\Real{
  \cosh (q-2|\mu|)\tau/2\,\,\cosh\tau/2\over\tau\,\sinh q\tau/2\,\sinh^{d-1}\tau/2}\cr
  }
  \eql{fullz3}
  $$
for $\tau=x+iy$ with $y<2\pi/q$. The integral converges in the relevant range of $\mu$. I
just compute it numerically.

In order to compare with results in [\pref{B}], figure 1 shows  the ratio,
$\caA_{n,d}(\mu)/\caA_{n,d}(0)$, of effective actions plotted against chemical potential
in three dimensions. Figure 2 has the results for $d=5$.

\epsfxsize=5truein \epsfbox{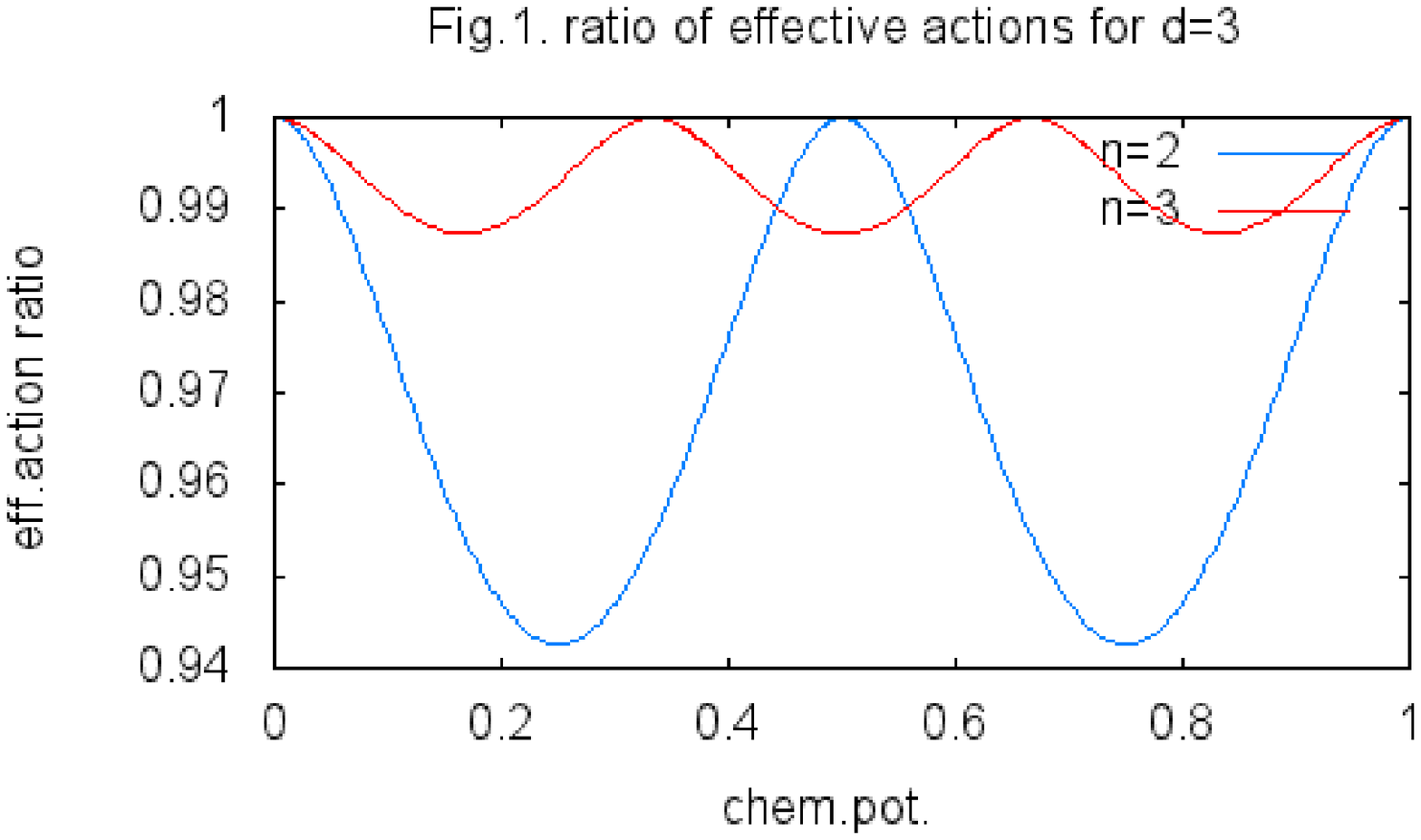}
\newpage

\epsfxsize=5truein \epsfbox{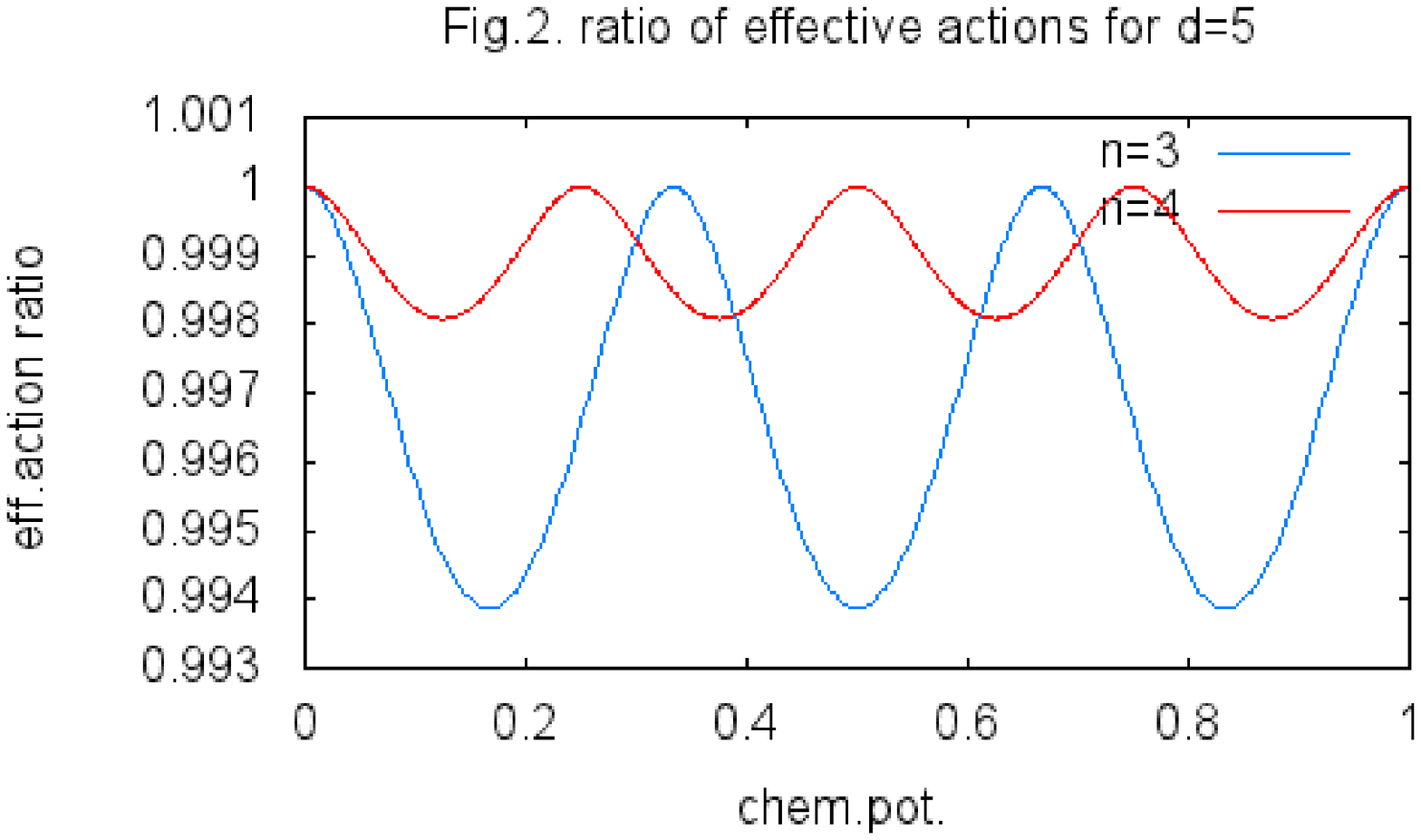}

\epsfxsize=5truein \epsfbox{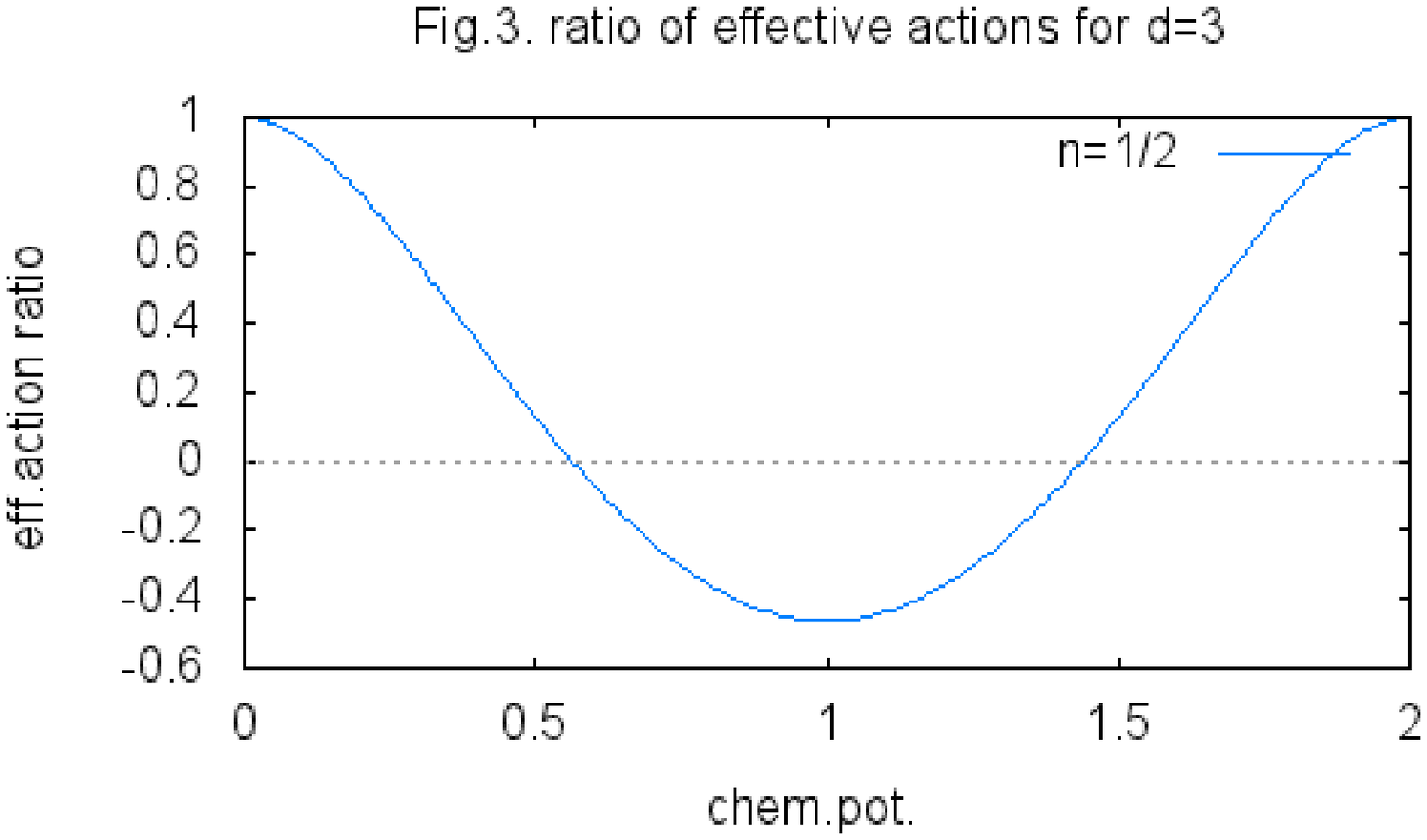}

\section{\bf3. Images}

An image formula given in [\pref{Dowcascone}], and further discussed in
[\pref{dowtwist}], has the  particular case, when expressed in terms of the integer
covering effective action (free energy), $\caA_n(\mu)$,
  $$
  \caA_n(\mu)=\sum_{s=0}^{n-1}\,\caA_1\big(\mu\pm{s\over n}\big)\,,\quad n\in\oN\,,
  \eql{imagef}
  $$
which can be used to check the numerical work or to obtain the left--hand side from
quantities on the ordinary sphere (with a flux). It could be termed a replica relation. The
equivalence of the $\pm$ signs is  a consequence of periodicity.

It is interesting to spell this out explicitly, the easiest case being dimension three (the
only odd one considered in [\pref{B}]). For this once, I use the explicit degeneracies.

Things simplify on the full sphere, $q=1=n$. I define, just for algebraic convenience, the
\zf\ occurring in (\peq{zf1}) in three dimensions,
  $$
  \ze(s,a)=\sum_{\bf m=0}^\infty{1\over
 \big( (a+{\bf1}{\bf . m})^2-{1\over4}\big)^s}=
 \sum_{m=0}^\infty\comb{m+2}{2}{1\over
 \big( (a+m)^2-{1\over4}\big)^s}\,.
 \eql{zeta}
  $$

In odd dimensions there is no multiplicative anomaly, and, factorising the eigenvalue, we
reach the auxiliary, or surrogate, \zf,
  $$
 \ze_A(s,w)=  {1\over2}\sum_{m=0}^\infty{(m+1)(m+2)\over(w+m)^s}\,,
  \eql{zf2}
  $$
where $w=a\pm1/2$ and $a$ takes the two values, according to (\peq{zf1}),
$a=1+|\mu|,\quad $ and $\quad a=2-|\mu| $.  Hence, to spell things out, defining
$\ol\mu=|\mu|-1/2$, \footnote{ I emphasise that $\mu$, here and later, has been
periodised as mentioned before.}
$w$ has the four values
  $$\eqalign{
  &w_p^{(\pm)}=p\pm\ol\mu\,,\quad p=1,2\,,
  }
  $$
showing the symmetry under $\ol\mu\to-\ol\mu$ so that one need calculate only for the
range $0<\mu <1/2$.

As is usual in these circumstances the degeneracy is expanded \footnote{ This is
employed in [\pref{B}].},
  $$
  (n+1)(n+2)=(w+n)^2+e(w+n)+f\,.
  $$
Hence, $e=3-2w$ and $f=2-3w+w^2$.

The auxiliary \zf, (\peq{zf2}) then becomes
  $$
  {1\over2}\big(\ze_R(s-2,w)+e\,\ze_R(s-1,w)+f\,\ze_R(s,w)\big)
  \eql{szed}
  $$
in terms of the Hurwitz \zf, as in many earlier works.

The theorem, [\pref{Dowcmp}],  is that the derivative at $s=0$ of (\peq{zeta}) is
  $$\eqalign{
  \ze'(0,a)&={1\over2}\big(\ze'_R(-2,w)+e(w)\,\ze'_R(-1,w)+f(w)\,\ze'_R(0,w)\big)\,.\cr
  }
  \eql{szed}
  $$

This is the same calculation as in [\pref{B}], only the mode structure and metric are
different.

An easy case is $\mu=1/2$ for which the coefficients have the values,
$$\eqalign{
  w&= \,\,\,2\,\,,\quad 1\,,\quad 2\,\,\,\,,\quad 1\cr
  e&= -1\,,\quad 1\,,\quad -1\,,\quad 1\cr
  f&=\,\,\,\,0\,,\,\quad0\,,\,\,\quad0\,,\quad0\,,
  }
  $$
and the corresponding total auxiliary \zf, (\peq{szed}), is,
  $$\eqalign{
  &\ze_R(s-2,2)-\,\ze_R(s-1,2)+\ze_R(s-2,1)+\,\ze_R(s-1,1)\cr
  &=\ze_R(s-2)-\,\ze_R(s-1)+\ze_R(s-2)+\,\ze_R(s-1)\cr
  &=2\ze_R(s-2)
  }
  $$
where $\ze_R(s)$ is the Riemann \zf.

Hence for (\peq{zf1})
  $$
  \caA_1(1/2)=\ze'(0,1/2)=4\ze'(-2)={\ze(3)\over2\pi^2}\,.
  $$

The values of $\caA_n(0)$ are old and standard. For example
$\caA_1(0)={\log2\over4}-{3\ze(3)\over{8\pi^2}}$ and
$\caA_2(0)={\log2\over4}+{\ze(3)\over8\pi^2}$. Therefore,
  $$
  \caA_2(0)=\caA_1(0)+\caA_1(1/2)\,,
  $$
which is the simplest example of the image relation, (\peq{imagef}). It can also be
derived from the integral (\peq{fullz3}) using (hyperbolic) trigonometric relations. Further
remarks can be found in section 4.

To work from the right--hand side of (\peq{imagef}) to the left, the values of $\caA_1(w)$
are required and  so I return to the expression (\peq{szed}) which is treated numerically.
The total \zf\ derivative is
  $$\eqalign{
  \ze'(0,\mu)&={1\over2}\sum_{p,\pm}\bigg(\ze'_R(-2,
  w_p^{(\pm)})+e(w_p^{(\pm)})\,\ze'_R(-1,w_p^{(\pm)})
  +f(w_p^{(\pm)})\,\ze'_R(0,w_p^{(\pm)})\bigg)\cr
  }
  \eql{szedd}
  $$

For the final term the standard result,
  $$
  \ze'_R(0,w)=\log(\Ga(w)/\sqrt{2\pi}\,.
  $$
can be used. The other two derivatives can also be related to Gamma--type functions but
it is probably best to treat them numerically, for example from asymptotic expansions
(\eg\ [{\pref{Elizalde}]). Of course if just a number is wanted, the integral form,
(\peq{fullz3}), is far easier, but alternative methods are handy checks and the two
methods do agree. This is the end of this specific example of images.

Although the image relation, (\peq{imagef}), is fairly general, in the present spherical
case it follows as a result of a `transformation' property of the \zf, (\peq{zf1}), which is,
written symbolically,

  $$
  \ze\big(s,\mu,q\mid{\bom\over n}\big)=
  \sum_{t=0}^{n-1}\ze\big(s,\mu+\sum{t\bom\over n},q\mid\bom\big)\,.
  \eql{trans}
  $$
$\bom/n$ stands for the set of $d$-degrees $\big(\om_1/n_1,\ldots,\om_d/n_d\big)$ and
  $$\eqalign{
  \sum {t\,\bom\over n}&\equiv{t_1\,\om_1\over n_1}+\ldots+{t_d\,\om_d\over n_d}\,.\cr
  }
  \eql{defs}
  $$

I give the expression in generality as it might be useful for later investigations. For
present application, $\bom={\bf1}$ and all the $n_i$ are equal to $1$, except $n_1\equiv
n$, the covering integer, so that all the $t_i$ are zero except $t_1\equiv t$. There is,
therefore, only one term in the sum in (\peq{defs}).

The image relation, (\peq{trans}), is derived by Barnes, [\pref{Barnesa}], for the \zf\
  $$
  \ze_d(s,a\mid\bom)
  =\sum_{{\bf {m}}={\bf 0}}^\infty{1\over(a+{\bf m.\bom})^s},\qquad
  \Real\, s>d\,,
 \eql{barn}
  $$
by the simple expedient of introducing residue classes mod ${\bf n}$. This can be applied
to (\peq{zf1}) and the sum in (\peq{trans}) is over these classes.

\section{\bf 4. Formal expressions. The Barnes Gamma function }

In [\pref{Dowcmp}] the auxiliary \zf\ method was derived and applied to general \zfs\ of
the type (\peq{zf1}). Factorising the denominators in (\peq{zf1}) leads to a sum of
Barnes \zfs\ (\peq{barn}), with four values for the parameter $a$,
$$\eqalign{
  a_p^{(\pm)}&={d+q\over2}\pm\ol\mu-p\cr
  &=b\pm\ol\mu-p\,,\quad p=0, 1\,,
  }
$$
so that $a_1^{(\pm)}=a_0^{(\pm)}-1$.

 The auxiliary \zf\ is then,
  $$
  \sum_{\pm}\sum_{p=0,1}\ze_d(s,a_p^{(\pm)}\mid\bom)\,,
  $$
where the $d$-degrees are $\bom=(q,\bf1)$, and the derivative of the original \zf,
(\peq{zf1}), reads,
  $$\eqalign{
  {1\over2}\ze'(0, \mu,q\mid\bom)&=\sum_{\pm}\sum_{p=0,1}\ze'_d(0,a_p^{(\pm)}\mid\bom)\cr
  &=\log{1\over\rho^4(\bom)}\prod_{\pm}\prod_{p=0,1}
  \Ga_{d}(a_p^{(\pm)}\mid\bom)\cr
  &=\log\prod_{\pm}{\Ga_{d+1}(a_0^{(\pm)}\mid\bom)\,\Ga_{d+1}(a_1^{(\pm)}\mid\bom)
  \over\Ga_{d+1}(a_0^{(\pm)}+1\mid\bom)\,
  \Ga_{d+1}(a_1^{(\pm)}+1\mid\bom)}\cr
  &=\log{
  \Ga_{d+1}(a_0^{(+)}-1\mid\bom)\Ga_{d+1}(a_0^{(-)}-1\mid\bom)
  \over\Ga_{d+1}(a_0^{(+)}+1\mid\bom)\Ga_{d+1}(a_0^{(-)}+1\mid\bom)}\,.\cr
  }
  \eql{gammaf}
  $$

In general, this presents no advantage for actual computation. However for the full sphere,
($q=1$), it can be taken further to elementary functions.

Dropping the dependence on $\bom={\bf1}$,
  $$\eqalign{
  {1\over2}\ze'(0, \mu,1)&=\log{\Ga_{d+1}(b+\ol\mu-1)\over\Ga_{d+1}(b-\ol\mu+1)}-
\log{\Ga_{d+1}(b+\ol\mu+1)
  \over\Ga_{d+1}(b-\ol\mu-1)}\,,\cr
  }
  \eql{earatio}
  $$
which can be evaluated in terms of standard functions if $d$ is odd. The details are given
in [\pref{DowGJMS}] and [\pref{Dowmasssphere}] with the result,
  $$\eqalign{
  {1\over2}\ze'(0,\mu,1)&=\log\,{\rm Sin}_{d+1}(b-\ol\mu+1)-\log\,{\rm Sin}_{d+1}(b-\ol\mu-1)\cr
  &=-{1\over d!}\int_{b-\ol\mu-1}^{b-\ol\mu+1}dz\,B^{(d+1)}_d(z)\,
  \pi\,\cot\pi z\,.\cr
  }
  \eql{int}
  $$
${\rm Sin}_d$ is Kurokawa's generalised sine function and the Bernoulli polynomial has
the product form,\mgn{do further manipulation --$>$ D's}
  $$
  B^{(d-1)}_d(x)=(x-1)(x-2)\ldots(x-d)\,.
  $$

To recapitulate, (\peq{int}) gives another, more analytical, formula for the effective action
(on a {\it full} sphere with flux) again showing the explicit dependence  on  $\mu$. For
example, the $\mu$ derivative could be found analytically as a function of $\mu$.
\footnote{Alternatively, this could be derived directly from the Barnes \zfs\ and then
integrated to give (\peq{int}).}

\begin{ignore}
  $$\eqalign{
 \pa_{\mu} \ze'(0,\mu,1)=&{1\over d!}\big(B^{(d+1)}_d(b-\ol\mu+1)
 -B^{(d+1)}_d(b-\ol\mu-1)\big)\,\pi\cot\pi\ol\mu\cr
  &={1\over(d-1)!}\,B^{(d)}_{d-1}(b+\ol\mu)\,\pi\cot\pi\ol\mu\cr
  &={1\over(d-1)!}\,\ol\mu^2\ldots\big(\ol\mu^2-(d-2)^2/4\big)\,\pi\,\cot\pi\ol\mu\,.\cr
  }
  \eql{diff}
  $$
\end{ignore}

Equation (\peq{int}) provides another means of numerically computing the effective action
on the full sphere and the values agree with the previous methods. Unfortunately, it again
applies only for odd dimensions. In the next section, I outline yet one more expression for
the effective action which holds for any dimension but involves a transcendental function.

Although these relations appertain just to the full sphere, they can be extended by images
to integer coverings.

\section {\bf 5. Another Gamma function form}

The following development can be found detailed in [\pref{DowGJMS}] (with references)
and makes use of Barnes' theory of the multiple $\psi$--function, $\psi^{(p)}_d(z)$ which
is the $p$th logarithmic derivative of the multiple $\Ga$--function, $\Ga_d$. Actually, the
only one required is $\psi_d^{(1)}(z)\equiv \psi_d(z)$ since by integration, trivially, the
desired ratio in (\peq{earatio}) is,
  $$\eqalign{
  \log {\Ga_d(z_2)\over\Ga_d(z_1)}&=\int_{z_1}^{z_2}\,\psi_d(z)\cr
  &={(-1)^{d-1}\over(d-1)!}\int_{z_1}^{z_2}dz\,\big(B^{(d)}_{d-1}(z)\,\psi(z)+
  Q_d(z)\big)
  }
  $$
where $\psi(z)$ is the ordinary $\psi$--function, if $\bom={\bf 1}$, and $Q$ is an easily
obtained polynomial. This equation results from recursion relations.

Since $\psi$  is numerically available, the effective action can be calculated this way for
even, as well as for odd, dimensions. I do not pursue this at this time.

\section{\bf 6. Charged R\'enyi entropy}

The definition of the R\'enyi entropy is,
  $$
   S_(\mu)={n\caA_d(1,\mu)-\caA_d(1/n,\mu)\over1-n}\,,\quad n=1/q\,,
   \eql{renyi}
  $$
where $\caA_d(q,\mu)$ is the effective action on the $d$--dimensional space--time
deformed by a conical singularity of angle $2\pi/q$ computed in the previous sections.

Figures 4 and 5 show the variation of the R\'enyi entropy against $n$, considered as a
continuous variable for various chemical potentials.

\epsfxsize=5truein \epsfbox{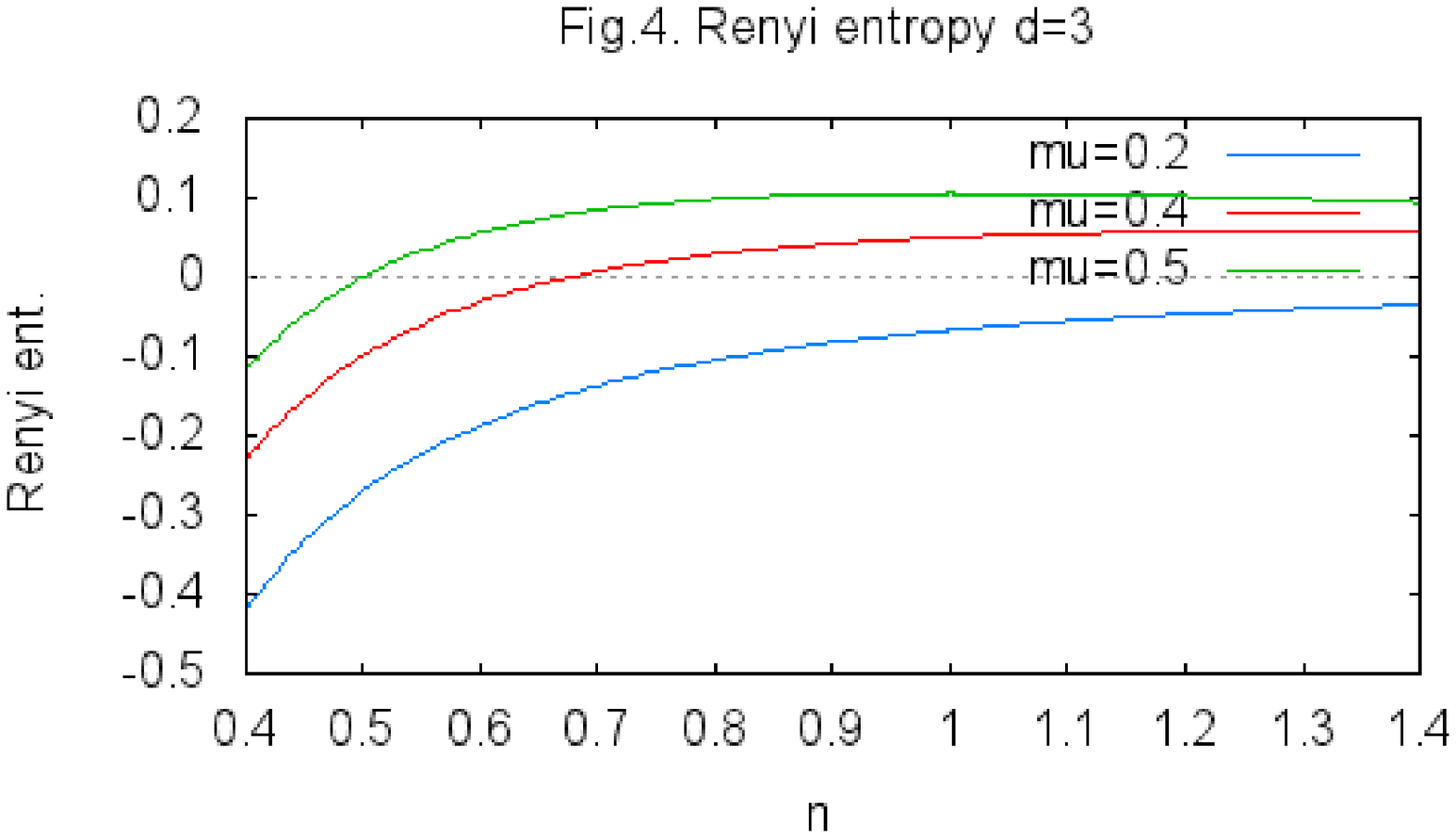}

\epsfxsize=5truein \epsfbox{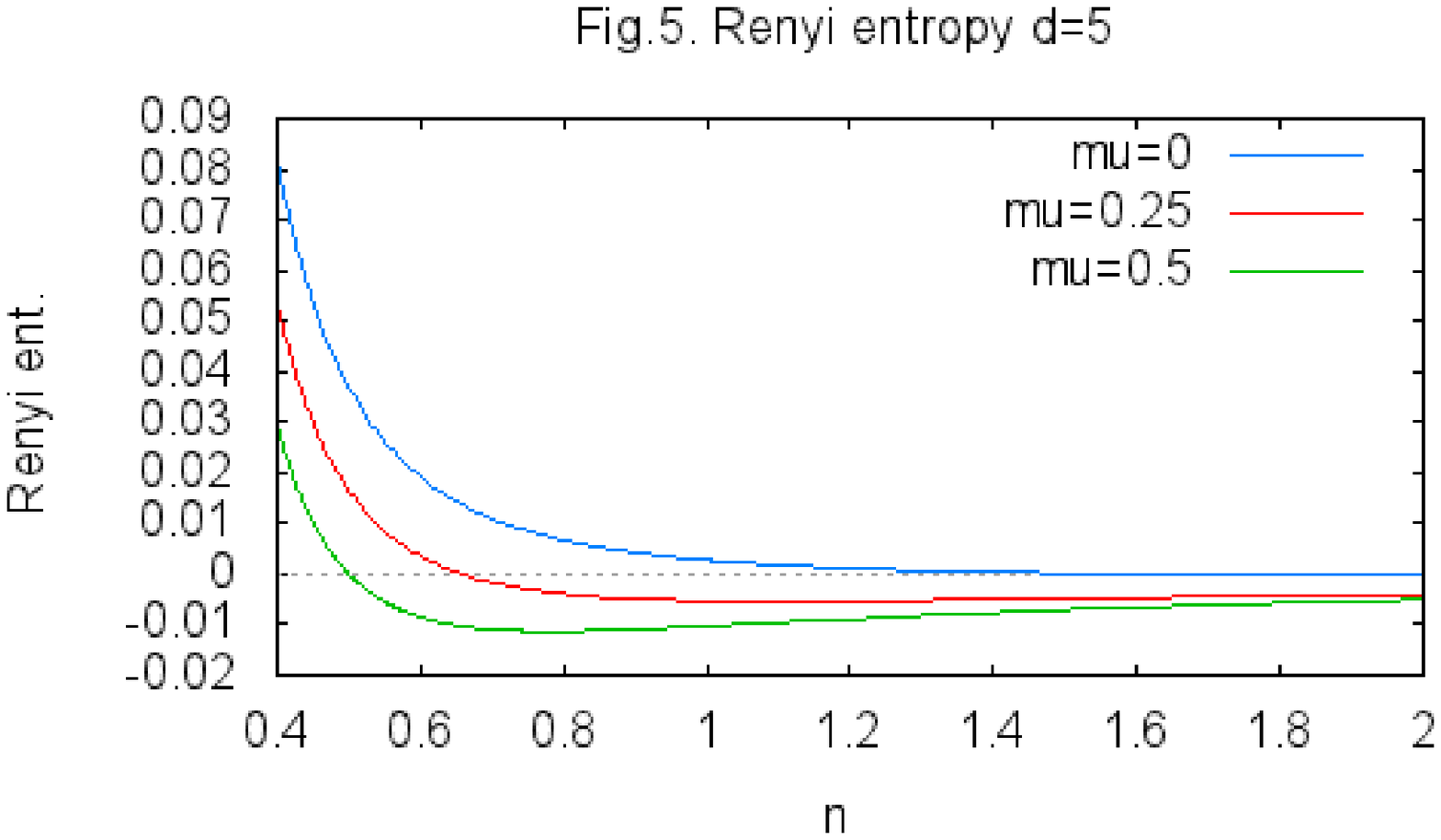}

\section{\bf 7. The $n\to\infty$ limit}

For large $n$, the entropy tends to a constant,
  $$
   \lim_{n\to\infty}S_n(\mu)={\lim_{q\to0}q\caA_d(q,\mu)}-\caA_d(1,\mu)
   \,,\quad n=1/q\,.
   \eql{renyi2}
  $$
The last term is the full sphere value and can be calculated by the other methods
described. There is an explicit integral for the first term and I consider it further, in its own
right, by a contour method which leads to a simple closed form.

I use $q$ in preference to $n$. The first thing to note is that, because of periodicity
$(\propto q)$, the limit is independent of $\mu$ \footnote{ A hint of this behaviour can be
seen in Figs. 1 and 2. The ripples become smaller and closer, the mean tending to one.}
(which is here defined to remain finite as $q\to0$). All values $\mu$ are equivalent to
$\mu=0$ giving, as the limit of (\peq{fullz3}),
  $$\eqalign{
  q\,\lim_{q\to0}\caA_d(q,\mu)
  &=-{1\over2^{d-1}}\int_{-\infty+iy}^{\infty+iy} dz\,
  {\cosh z\over z^2\,\sinh^{d-1}z}\cr
  &={1\over2^{d-2}(d-2)}\int_{-\infty+iy}^{\infty+iy} d\tau\,
  {1\over z^3}
  {1\over\sinh^{d-2}z}\cr
  &\equiv \caA_d^\infty\,.
  }
  \eql{fullzl}
  $$

\begin{ignore}
\section {[ Factors of 2  }

$1/2^{d-1}$ from contour form of eff. action. Factor of 2 from limit of $\sinh(q\tau/2)$ in
denominator as $q\to0$. Factor of $1/2$ from change $2\tau=z$. Factor of 2 from partial
integration (first differentiation ) . Total:
  $$
  {1\over 2^{d-1}}\times 2 \times {1\over2}\times2={1\over2^{d-2}}
  $$
Then there will be another 2 from residues making $1/2^{d-3}$.]

\end{ignore}

It is possible to carry the calculation forward for any (odd) dimension. For this purpose, as
in earlier works, I employ the expansion,
$$
    \cosech^{2r+1}z={(-1)^r\over(2r)!}\sum_{\rho=0}^r(-1)^\rho\,\caG^r_\rho \,
    {d^{2\rho}\over dz^{2\rho}}\,\cosech\, z\,,
    \eql{sechs}
  $$
for odd powers in terms of even derivatives and then integrate by parts. The integer
coefficients, $\caG^r_\rho $, are known. This results in, \mgn{factor of 2 on top from
residues}
$$\eqalign{
  \caA_d^\infty&={2(-1)^r\over2^{2r+1}(2r+1)!}\sum_{l=1}^\infty
  \sum_{\rho=0}^r{(2+2\rho)!\over 2!}\,{\caG_\rho^r\over\pi^{2\rho+2}}
\sum_{l=1}^\infty(-1)^l{1\over l^{2\rho+3}}\cr
&=-{(-1)^r\over2^{2r+1}(2r+1)!}\sum_{\rho=0}^r(2+2\rho)!\,\caG_\rho^r\,
{\eta(2\rho+3)\over\pi^{2\rho+2}}\,,
}
  \eql{fullzl1}
  $$
where $d=2r+3$.

The numbers, $\caG_\rho^r$, have been tabulated. Some are below,
  $$
\matrix{r&0&1&2&3&4&\rho\cr{}&1&1&9&225&11025&0\cr {}&{}&1&10&259&12916&1
\cr{}&{}&{}&1&35&1974&2\cr{}&{}&{}&{}&1&84&3\cr{}&{}&{}&{}&{}&1&{\,\,4}\,,
}
$$
giving for example,
  $$\eqalign{
\caA_3^\infty&=
{\eta(3)\over\pi^{2}}\approx0.091345371175179\cr
\caA_5^\infty&=
-{\eta(3)\over24\pi^{2}}-{\eta(5)\over 2\pi^4}\approx-0.0087959392887171\cr
\caA_7^\infty&=
{3\eta(3)\over640\pi^{2}}+{\eta(5)\over 16\pi^4}+{3\eta(7)\over16\pi^6}
\approx 0.0012455025438325\,,\cr
}
\eql{values}
  $$
the first of which is given by Klebanov {\it el al}, [\pref{KPSS}].

Purely numerically, the integral form (\peq{fullzl}) is more convenient and, although the
integral is suppose to hold only for $d$ integral and odd, in figure 8  the limiting value is
plotted against a (continuous) dimension. This interpolates the values, (\peq{values}).

\epsfxsize=5truein \epsfbox{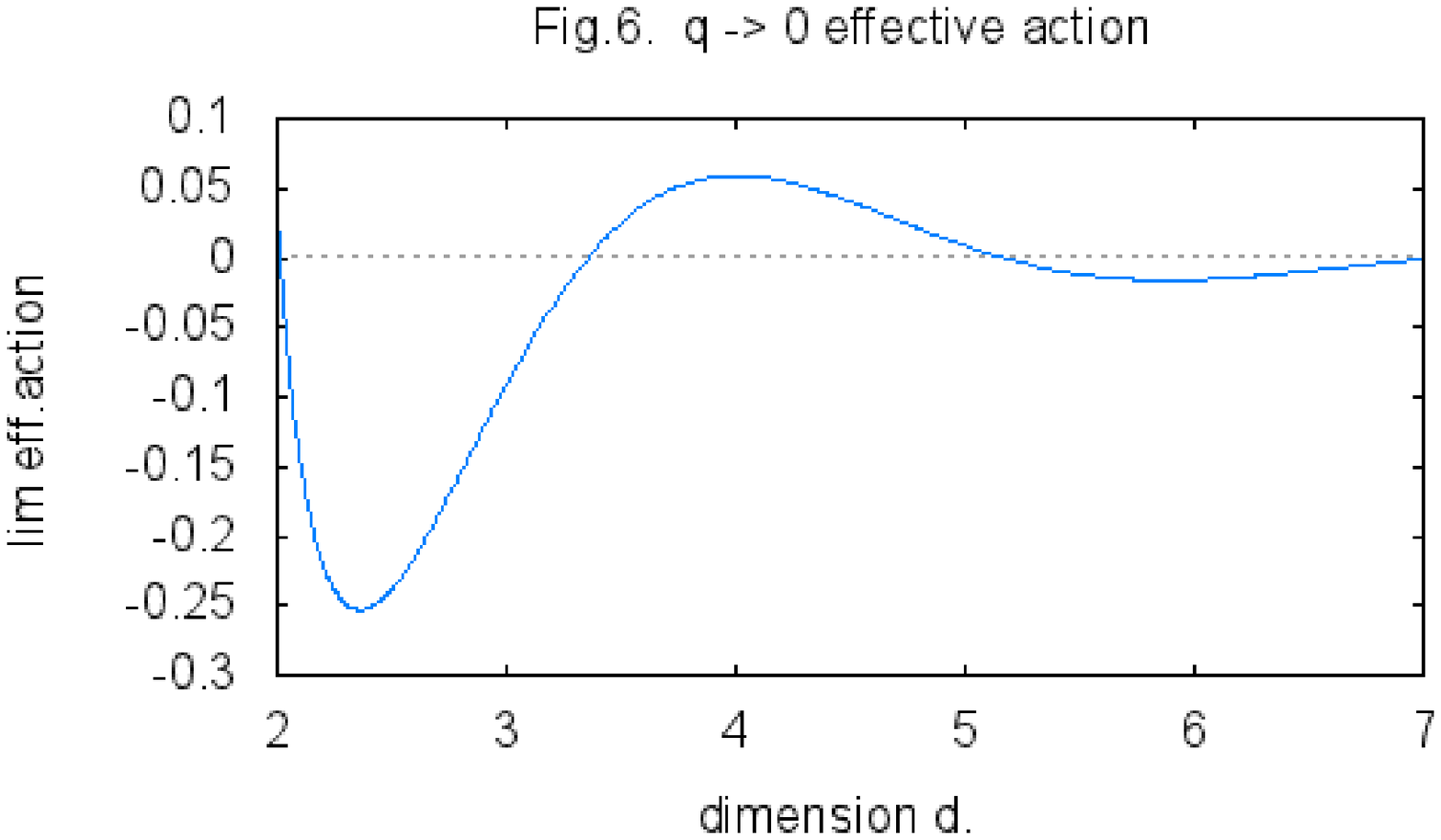}

For zero chemical potential, Bueno, Myers and Witczak--Krempa, [\pref{BMW}], have
discussed the relation between a corner coefficient, $\si^n_d$, and the effective action,
$\caA_d(1/n,0)$, in the $n\to\infty$ limit. $\si^n_d$ is related to the conformal
dimension, $h^n_d$, of twist operators and so I investigate the connection between
$h^\infty$ and $\caA_d^\infty$, for odd dimensions. In fact this relation holds, rather
trivially, for any (Euclidean) chemical potential because, in the limit, they are all
equivalent to zero as a consequence of periodicity.

In [\pref{dowtwist}], using a conformal transformation to  a conical space, an expression
for $h^n_d$ was derived from the energy density. As shown there, this depends on the
integral, (I have already set the chemical potential to zero),
  $$
  W_d(q)=\sin\pi q\int_0^\infty {d\tau\over\cosh^d \tau/2}\,
  {1\over \cosh q\tau-\cos q\pi}\,,
  \eql{wd}
  $$
by, for conformal scalars,
  $$
  h^n_d=-{2\Ga(d/2)\over2^d\pi^{d/2}}\,\bigg(W_d(q)-{d-2\over d-1}\,W_{d-2}\bigg)\,.
  \eql{hconf}
  $$

Taking the $n\to\infty$ limit of (\peq{wd}),
  $$\eqalign{
  \lim_{q\to0}W_d(q)&={2\over q}\int_0^\infty {d\tau\over\cosh^d \tau/2}\,
  {\pi\over \tau^2+\pi^2}\cr
  &\equiv {1\over q}\,W_d^\infty\,.
  }
  \eql{jd}
  $$
I likewise define $h^\infty_d={1\over n}\,\lim_{n\to \infty}\,h^n_d$.

The integral in (\peq{jd}) was encountered, and computed,  in [\pref{Dowren}] in
connection with the effective action on a {\it full} sphere (\ie $q=1$). It takes the form of
a sum of $\eta$--functions plus a $\log2$ term with coefficients involving the
$\caG^r_\rho$. For completeness, I briefly recall the results.

In terms of the notation in [\pref{Dowren}],
  $$
  W_d^\infty=2J(d)\,.
  $$

To calculate (\peq{jd}), this time one uses the recursion,
  $$
    \sech^{2r+1}x={1\over(2r)!}\sum_{\rho=0}^r(-1)^\rho\,\caG^r_\rho \,
    {d^{2\rho}\over dx^{2\rho}}\,\sech\, x\,,
    \eql{sechs1}
  $$
and $J(d)$ is found to be (now $d=2r+1$),
  $$\eqalign{
   J(2r+1)
   &=\sum_{\rho=1}^r \,\caG_\rho^r\,{ (2\rho)!\over(2r)!}
   \,{\eta(2\rho+1)\over\pi^{2\rho}}+{1\over(2r)!}\caG^r_0\,\log2\,.
   }
  $$

On taking the conformal combination, $J(2r+1)-{2r-1\over2r}\,J(2r-1)$, one finds, firstly,
that the $\log2$ terms cancel in view of the recursion relation
  $$
    \caG^r_\rho=(2r-1)^2\,\caG^{r-1}_\rho+\caG^{r-1}_{\rho-1}\,,
    \eql{erecur}
  $$
and the rest of the series equals,
   $$
\sum_{\rho=1}^r \,\caG_\rho^r\,{ (2\rho)!\over(2r)!}
   \,{\eta(2\rho+1)\over\pi^{2\rho}}-{2r-1\over2r}
   \sum_{\rho=1}^{r-1} \,\caG_\rho^{r-1}\,{ (2\rho)!\over(2r-2)!}
   \,{\eta(2\rho+1)\over\pi^{2\rho}}\,.
   $$
Extend the second sum to $\rho=r$ which is possible because $\caG_\rho^r$ is zero if
$\rho>r$. Then one finds for this combination,
  $$\eqalign{
   &\sum_{\rho=1}^r (2\rho)!\bigg[\,\caG_\rho^r\,{ 1\over(2r-1)}
   \,-(2r-1)
   \,\caG_\rho^{r-1}\,
   \,\bigg]{\eta(2\rho+1)\over\pi^{2\rho}}{1\over2r(2r-2)!}\cr
   &=\sum_{\si=0}^{r -1}\,\caG_{\si}^{r-1}{(2\si+2)!\over(2r)!}
   \,{\eta(2\si+3)\over\pi^{2\rho}}\,.\cr
   }
   $$

A comparison with (\peq{fullzl1}) shows agreement, to a factor, if one remembers that in
(\peq{fullzl1}), $d=2r+3$. To allow for this,  shift $r$ by 1, then

$$\eqalign{
\caA_d^\infty&={(-1)^r\over2^{2r-1}(2r-1)!}
\sum_{\rho=0}^{r-1}(2+2\rho)!\,\caG_\rho^{r-1}\,
{\eta(2\rho+3)\over\pi^{2\rho+2}}\,,\cr
}
  \eql{fullzl2}
  $$
and, from (\peq{hconf}),
$$
  h_d^\infty=-{4\Ga(r+1/2)\over2^{2r+1}\,\pi^{r+1/2}\,(2r)!}
  \sum_{\si=0}^{r -1}\,\caG_{\si}^{r-1}(2\si+2)!
   \,{\eta(2\si+3)\over\pi^{2\rho}}\,.
  \eql{hconf2}
  $$

So I find the relation,
  $$\eqalign{
h_d^\infty&=-(-1)^r(2r-1)!2^{2r-1}\,
{4\Ga(r+1/2)\over2^{2r+1}\pi^{r+1/2}\,(2r)!}\caA_d^\infty\cr
&=(-1)^{r-1}\,
{\Ga(r+1/2)\over2\pi^{r+1/2}\,r}\,\caA_d^\infty\,,
}
\eql{hdinf}
  $$
and, in particular,
  $$
h_3^\infty={1\over4\pi}\caA_3^\infty\,.
\eql{h3inf}
  $$

According to [\pref{BMW}] the limit, $\si^\infty_3$, of the corner coefficient, $\si^n_3$,
is related to $h_3^\infty$ by $\si^\infty_3=h_3^\infty/\pi$ (in three dimensions), and so
(\peq{h3inf}) implies that
  $$
  \si^\infty_3={1\over4\pi^2}\caA_3^\infty\,.
  \eql{sig3inf}
  $$

I have thus confirmed, in a rather detailed fashion, the results of  [\pref{BMW}] who
prove this relation for {\it any} 3d conformal field theory using the mapping to the
hyperbolic cylinder. The result was found earlier by Elvang and Hadjiantonis,
[\pref{EandH}], for free bosons and fermions, from calculated expressions for the
$\si^n_3$.

In [\pref{BandM}] a conjecture is made regarding the relation between the corner
coefficient and the conformal weight in any dimension. Assuming this relation, a
generalisation of (\peq{sig3inf}) can be found, for odd dimensions $d=2r+1$,\footnote{ A
similar conclusion holds for even $d$ using the explicit polynomial expression for $h^n_d$.
The details might be presented at a later time.}
   $$
   \si_d^\infty=(-1)^{r-1}{2^{2r-4}\over\pi^2}\bigg[{(r-1)!\over(2r-1)!!}\bigg]^2
   \caA_d^\infty\,.
   \eql{sigdinf}
   $$

As mentioned above, the integral $J(d)$ in (\peq{jd}) occurs in the effective action on the
full sphere.  I give the expression which is, [\pref{Dowren}],
  $$
  \caA_{2r+1}(1,0)={(-1)^r\over 2^{2r-1}}\big(J(2r+1)-J(2r-1)\big)\,,
  $$
which shows that the the limiting form of the R\'enyi entropy, (\peq{renyi2}), can be
written just in terms of the basic integral, $J(d)$.

\section{\bf 8. R\'enyi entropies on even spheres revisited}

The R\'enyi entropy, or, rather, just its universal log coefficient (at zero chemical
potential) was computed on even spheres, dimension by dimension, some time ago by
Casini and Huerta, [\pref{CaandH}], using their mapping to the hyperbolic cylinder (see
also [\pref{KPSS}]). In [\pref{Dowren}] the same quantity was found from the
heat--kernel via known expressions for the conformal anomaly for any (even) dimension.
All this for scalar fields, to which I am always restricting myself. The answer was given in
terms of generalised Bernoulli polynomials, which, although easily obtained, are, perhaps,
not conveniently organised for some purposes. In this section I present an alternative
arrangement which offers a simplification and extension of known results.

Rather than extracting the log coefficient using a heat--kernel expansion, of one sort or
another, I employ the method described in [\pref{doweven}] which involves off--shell
thermodynamic arguments. The logarithm arises from the divergent integration up to the
de Sitter horizon, as opposed to a regularised odd hyperbolic volume or a generic UV cut
off.

The R\'enyi entropy is determined, (\peq{renyi}), by the effective action, $\caA$. The
thermodynamic free--energy, $F$, is obtained from $\caA$ by removing the (imaginary)
time circle factor,
  $$
  \caA\big({q}\big)=-\beta\,F=-\int^\beta d\beta\,E(\beta)\,,\quad q=2\pi/\be\,,
  \eql{effact}
  $$
where I have introduced the internal energy, $E(\beta)$. The entropy is then just
(\peq{renyi}) for $\mu=0$,
   $$
   S_q={\caA_d(1)-q\caA_d(q)\over q-1}\,.
   \eql{renyi3}
  $$

$E(\beta)$ is found from the integral of the energy density $\av {T_{00}}_{dS}$ on de
Sitter, which is given by conformal transformation from that on a conical flat space, derived
somewhat earlier. The general structure is,
  $$
  \av {T_{00}}_{dS}={1\over (4\pi)^{d/2}}\,P(d,q)\,{(1+Z^2)^d\over2^d Z^d}\,.
  \eql{t00de}
  $$
$P$ is a polynomial in $q$ determined by the field dynamics. The last factor is a geometric
one giving the dependence on the radial--like coordinate, $Z$, of the static de Sitter
metric,
  $$
  ds^2={4a^2\over(1+Z^2)^2}\,\big(Z^2d(t/a)^2-dZ^2\big)
  -a^2\bigg({1-Z^2\over1+Z^2}\bigg)
  ds^2\big|_{(d-2)-sphere}\,.
  \eql{desmet}
  $$

The total energy, $E$, is obtained by integrating (\peq{t00de}) over the spatial section of
(\peq{desmet}) up to an infinitesimal distance, $\ep$, from the horizon, $Z=0$. The
expansion in $\ep$ then allows the coefficient of $\log\ep$  to be extracted.

If the effective action is constructed via the thermodynamic relation, (\peq{effact}), with
the energy coming from (\peq{t00de}), the  log coefficient in $\caA(q)$ is found to be,
after algebra,
  $$\eqalign{
{\rm logcoeff} \caA(q)&=(-1)^{d/2}{1\over2^{d-2}\,\Ga(d/2)}\int^q dq\,{P(d,q)\over q^2}\cr
}
\eql{logcoeff2}
   $$
and the coefficient in the entropy follows immediately from (\peq{renyi3}) if the integral
can be found. For this one needs the polynomial, $P(d,q)$, which follows by conformal
transformation from the energy density on flat conical space (equivalent to Rindler space)
and gives,

$$\eqalign{
  \av{T_{00}}_{dS}&={(d-1)\Ga(d/2)\over(4\pi)^{d/2}}\,{q\over\pi}\,
  \bigg(W_d(q)-{d-2\over d-1}\,W_{d-2}(q)\bigg){(1+Z^2)^d\over2^d Z^d}\,.\cr
  }
  \eql{t00}
  $$

 Hence the polynomial is (I set $d=2g$),
   $$
   P(2g,q)=(2g-1)\Ga(g)\,{q\over\pi}\,
  \bigg(W_{2g}(q)-{2(g-1)\over 2g-1}\,W_{2g-2}(q)\bigg)\,.
  \eql{pee}
   $$

So far the analysis is the same as in [\pref{doweven}]. There, the quantities, $W_d$,
were given as generalised Bernoulli polynomials and the integration in (\peq{logcoeff2})
performed dimension by dimension to give, it turns out, the conformal anomaly, in the
case of the entanglement entropy. The difference here is that they are computed via a
trigonometric sum, which has an image interpretation, derived by Jeffery, [\pref{Jeffery}],
see [\pref{dowtwist,Dowcosecs}], which allows the integration to be done explicitly for
{\it all} $d$.

Jeffery's expression is,
$$
   {q\over\pi}\,W_{2g}(q)={2^{2g-1}\over\Ga(2g)}\,\sum_{i=0}^{g-1}
   {\Ga(2i+2)\over 2^{2i}}\,A^g_i\,(1-q^{2i+2})\,{\ze(2i+2)\over\pi^{2i+2}}\,,
   \eql{jeff}
  $$
where the $A^g_i$ are constants related  to central factorial numbers. They vanish if $i\ge
g$ and were early tabulated. (See also [\pref{Dowrenexp}], and below.) \footnote{ The
validity of (\peq{jeff}) for {\it any} real $q$ depends on Carlson's theorem.
[\pref{dowtwist}] has a few more details. }

To get the energy density,  according to (\peq{t00}), a conformal combination is required,
  $$\eqalign{
  W_{2g}(q)-{2g-2\over2g-1}\,W_{2g-2}(q)
  ={\pi\over q}{\,2^{2g-1}\over\Ga(2g)}\sum_{i=0}^{g-1}
   {\Ga(2i+2)\over 2^{2i}}\,B^g_i\,\big(1-q^{2i+2}\big)\,{\ze(2i+2)\over\pi^{2i+2}}\,,
  }
  \eql{whyq}
  $$
using $A_{g-1}^{g-1}=0$ and where the $B^g_i$ are the easily found constants,
  $$
   B^g_i=A^g_i-(g-1)^2\big(A_i^{g-1}-\de_{i,g-1}\big)
   =A^{g-1}_{i-1}+(g-1)^2\de_{i,g-1}\,,
  $$
upon using the recursion, \eg\ [\pref{Steffensen}],
  $$
  A^{k+1}_{i+1}=A^k_i+k^2\,A^k_{i+1}\,.
  \eql{recurs}
  $$
The definition of $B$ is such that $B^k_i=A^{k-1}_{i-1}$, except $B^k_{k-1}=1$.

Then,
  $$\eqalign{
  P(2g,q)&={(2g-1)\Ga(g)\,\over\Ga(2g)}\sum_{i=1}^{g-1}2^{2g-2i-1}
   \Ga(2i+2)\,B^{g}_{i}\,\big(1-q^{2i+2}\big)\,{\ze(2i+2)\over\pi^{2i+2}}\cr
   }
  $$
employing $A^k_{k-1}=1$ and  $B^k_0=0$.\mgn{agrees with Phys Rev cone. See
AcoeffsThiele.wxm}

There is no term proportional to $q^2$. This is a consequence of conformal coupling,
[\pref{Dowcascone}], \ie of the specific combination in (\peq{pee}).

Then the integral in (\peq{logcoeff2}) is,
  $$\eqalign{
     \int^q dq\,&{P(2g,q)\over q^2}=\cr
     &-{(2g-1)\Ga(g)\,\over\Ga(2g)\,q}\sum_{i=1}^{g-2}2^{2g-2i-1}\,
   \Ga(2i+2)\,B^{g}_{i}\,\bigg(1+{q^{2i+2}\over2i+1}
   \bigg)\,{\ze(2i+2)\over\pi^{2i+2}}\,,\cr
   }
  $$
and, from (\peq{logcoeff2}),
  $$\eqalign{
{\rm logcoeff} \caA(q)
&={(-1)^{g}\over q(2g-2)!}\sum_{i=1}^{g-1}\,
   \Ga(2i+2)\,B^{g}_{i}\,\bigg(1+{q^{2i+2}\over2i+1}
   \bigg)\,{\ze(2i+2)\over2^{2i-1}\pi^{2i+2}}\,,\cr
}
\eql{logcoeff3}
   $$
with the particular on shell value,
  $$\eqalign{
   &{(-1)^{g}\over (2g-2)!}\sum_{i=1}^{g-1}\,
   (2i)!(i+1)\,B^{g}_{i}\,\,{\ze(2i+2)\over2^{2i-2}\pi^{2i+2}}\,.
}
\eql{logcoeff4}
   $$

The log coefficient in the R\'enyi entropy, (\peq{renyi3}), can then be assembled,
$$\eqalign{
  {\rm logcoeff}\,S_q=
   &={(-1)^{g}\over(2g-2)!}
   \sum_{i=1}^{g-1}\,
   (2i)!\,\,B^{g}_{i}\,\bigg({q^{2i+2}
   -1\over q-1}\bigg)\,{\ze(2i+2)\over2^{2i-1}\pi^{2i+2}}\,.\cr
}
\eql{renent}
  $$

Casini and Huerta, [\pref{CaandH}], obtain an expression of similar form, but the
constants are different as they come from the heat--kernel on hyperbolic space. A general
formula is not given, although the constants are calculated for specific dimensions.

The only unknowns in (\peq{renent}) are the constants, $A_i^g$. As mentioned, these are
tabulated or can easily be found from the recursion, (\peq{recurs}), so that
(\peq{renent}) can be quickly programmed.

One asset of the form, (\peq{renent}), is that it facilitates the limit $q\to1$ which gives
for the log coefficient of the entanglement entropy, $S_1$,
  $$\eqalign{
  {\rm logcoeff} S_1
   ={(-1)^{g}\over(2g-2)!}
   \sum_{i=1}^{g-1}\,
   (2i)!(i+1)\,\,B^{g}_{i}\,{\ze(2i+2)\over2^{2i-2}\pi^{2i+2}}\,.\cr
}
\eql{entent}
  $$
The same as (\peq{logcoeff4}).

The heat--kernel approach (or \zf) gives an expression for logcoeff $S_q$  in terms of
conformal anomalies, which reduces to just the standard round anomaly in the case of
$q=1$. (\peq{entent}) is therefore expected to equal this, as evaluation confirms.

Direct derivations of the anomaly via the \zf\ and expansion of the degeneracy,
[\pref{CandT}], also produce a sum over Riemann \zfs\ but it would need some
manipulation to show formal agreement with (\peq{entent}).

An even simpler form can be obtained by rewriting the generalised Bernoulli form,
[\pref{Dowhyp}], of the anomaly, $\caC$, to give (see [\pref{DowGJMS}], [\pref{Diaz}])
for complex fields,
  $$\eqalign{
  \caC_{2g}={4(-1)^{g}\over \,(2g)!}
  \int_0^1dt\,\prod_{i=0}^{g-1}\big(i^2-t^2\big)\,,\cr
  }
  $$
which can easily be evaluated dimension by dimension. However, use of the coefficients,
$A_i^g$, allows \footnote{ Relatedly, the coefficients occur when expanding the
degeneracy. The connection with the direct \zf\ and hyperbolic cylinder approaches could
then be made.} the integral to be done, rather trivially, for {\it all} $d$, yielding the
compact expression,
  $$\eqalign{
  \caC_{2g}&={4(-1)^{g}\over \,(2g)!}
  \sum_{j=1}^{g}(-1)^{j}\int_0^1dt\,t^{2j}\,A_{j-1}^g\cr
  &={4(-1)^{g}\over \,(2g)!}
  \sum_{j=1}^{g}(-1)^{j}\,{A_{j-1}^g\over2j+1}\,,\cr
  }
  \eql{confanom}
 $$
which allows hand calculation, for smallish dimensions, more readily than (\peq{entent}).

The equality of (\peq{entent}) and (\peq{confanom}) entails the curious sum rule
involving central factorials,
  $$
  \sum_{j=1}^{g}(-1)^{j}\,{A_{j-1}^g\over2j+1}=
  g(2g-1)
   \sum_{i=1}^{g-1}\,
   (2i)!(i+1)\,\,B^{g}_{i}\,{\ze(2i+2)\over2^{2i-1}\pi^{2i+2}}\,.
  $$

\begin{ignore}

\section{Tidy up [\pref{}]. Summarise work}

deS metric
  $$\eqalign{
  &(1-r^2)\,dt^2-{1\over 1-r^2}\,dr^2-r^2\,dS_{d-2}\cr
  &={4\over(1+Z^2)^2}(Z^2\,dt^2-dZ^2)-\bigg({1-Z^2\over1+Z^2}\bigg)^2
  dS_{d-2}
  }
  $$

Use $Z$
  $$
  Z^2={1-r\over1+r}\,,\quad r={1-Z^2\over1+Z^2}
  $$

  $$
  \sqrt {-g}= {4Z\over(1+Z^2)^2}\bigg({1-Z^2\over1+Z^2}\bigg)^{d-2}\sqrt{g_S}
  $$

  $$
  \av{T_0^0}\big|_{deS}={P(d,q)\over(4\pi)^{d/2}}{(1+Z^2)^d\over 2^d\,Z^d}
  $$

  $$\eqalign{
  E(\be)=\int\av{T_0^0}\big|_{deS}\,d\Si
  &={P(d,q)\over(4\pi)^{d/2}}\int{(1+Z^2)^d\over 2^d\,Z^d}
  \bigg({1-Z^2\over1+Z^2}\bigg)^{d-2}dZ\,\int d\Om_{d-2}\cr
  &={P(d,q)\over(4\pi)^{d/2}}\int{(1+Z^2)^d\over 2^d\,Z^d}
  {4Z\over(1+Z^2)^2}\bigg({1-Z^2\over1+Z^2}\bigg)^{d-2}
  dZ\,\int d\Om_{d-2}\cr
  &=|S^{d-2}|{P(d,q)\over(4\pi)^{d/2}}\int dZ\,{(1-Z^2)^{d-2}\over 2^{d-2}\,Z^{d-1}}
  \cr
  }
  $$

$d\Si$ = deS spatial vol. area.

Extracting the logcoeff, according to [\pref{doweven}]
  $$\eqalign{
  &{\rm logcoeff }\int\av{T_0^0}\big|_{deS}\,d\Si
  =(-1)^{d/2}|S^{d-2}|\comb{d-2}{d/2-1}{P(d,q)\over2^{d-2}(4\pi)^{d/2}}\cr
&=(-1)^{d/2}|S^{d-2}|\comb{d-2}{d/2-1}{P(d,q)\over2^{2d-2}\pi^{d/2}}\cr
&=(-1)^{d/2}{1\over2^{d-1}\pi\Ga(d/2)}P(d,q)\cr
  }
  $$

 From (\peq{effact})
   $$\eqalign{
\caA(2\pi/\be)&=(-1)^{d/2+1}{1\over2^{d-1}\pi\Ga(d/2)}\int^\be d\be\,P(d,q)\cr
\caA(q)&=(-1)^{d/2}{1\over2^{d-2}\,\Ga(d/2)}\int^q dq\,{P(d,q)\over q^2}\cr
}
   $$

\end{ignore}

\section{\bf 9. Comments}

The numerical methods used in this work have the significant advantage that the
degeneracies, which can become quite involved, do not appear and therefore need not be
expanded. The calculation of [\pref{B}] uses the explicit forms of these degeneracies
leading to a traditional sum of derivatives of Hurwitz \zfs, which have still to be
numerically evaluated. \footnote{ The degeneracies are coded into the expression
(\peq{fullz3}) and would emerge, in one form or another, on developing the integral using
a contour approach into \zfs.}

Another feature of the present technique is that it allows the covering, $n$, to be taken
continuous. It also allows an explicit treatment of the infinite $n$ limit.

 \vglue 20truept
 \noin{\bf References.} \vskip5truept
\begin{putreferences}
    \ref{Barnesa}{Barnes,E.W. {\it Trans. Camb. Phil. Soc.} {\bf 19} (1903)
  374.}
  \ref{Barnesb}{E.W.Barnes {\it Trans. Camb. Phil. Soc.} {\bf 19} (1903)
  426.}
  \ref{Elizalde}{Elizalde,E. {\it Math. of Comp.} {\bf 47} (1986) 347.}
  \ref{doweven}{Dowker,J.S. {\it Entanglement entropy on even spheres} ArXiv:1009.3854.}
  \ref{CandT}{Copeland,E. and Toms,D.J. \np {255}{1985}{201}.}
  \ref{Dowmasssphere}{Dowker,J.S. {\it Massive sphere determinants} ArXiv:1404.0986.}
  \ref{dowtwist}{Dowker,J.S. {\it Conformal weights of charged R\'enyi entropy
  twist operators for free scalars in arbitrary dimensions.} ArXiv:1508.02949.}
   \ref{Dowlensmatvec}{Dowker,J.S. {\it Lens space matter determinants in the vector
   model},  ArXiv:\break 1405.7646.}
   \ref{dowtwist}{Dowker,J.S. {\it Conformal weights of charged R\'enyi entropy twist
   operators for free scalar fields in arbitrary dimensions} ArXiv:1509.00782.}
   \ref{GandS}{Gel'fand,I.M. and Shilov,G.E. {\it Generalised Functions} Vol.1 (Academic Press,
   New York, 1964.}
   \ref{BandS}{Bogoliubov,N.N. and Shirkov,D.V. {\it Introduction to the theory of quantized
   fields} (Interscience, New York, 1959.).}
   \ref{dowsignch}{Dowker,J.S. \jpa{2}{1969}{267}.}
   \ref{MandD}{Dowker,J.S. and Mansour,T. {\it J.Geom. and Physics} {\bf 97} (2015) 51.}
   \ref{dowaustin}{Dowker,J.S. 1979 {\it Selected topics in topology and quantum
    field theory}
    \ref{Dowmultc}{Dowker,J.S. \jpa{5}{1972}{936}.}
    (Lectures at Center for Relativity, University of Texas, Austin).}
   \ref{Dowrenexp}{Dowker,J.S. {\it Expansion of R\'enyi entropy for free scalar fields}
   ArXiv:1408.0549.}
   \ref{EandH}{Elvang, H and  Hadjiantonis,M.  {\it Exact results for corner contributions to
   the entanglement entropy and R\'enyi entropies of free bosons and fermions in 3d} ArXiv:
   1506.06729 .}
   \ref{SandS}{T.Souradeep and V.Sahni \prD {46} {1992} {1616}.}
   \ref{CandH}{Casini H., and Huerta,M. \jpa{42}{2009}{504007}.}
   \ref{CandH2}{Casini H., and Huerta,M. J.Stat.Mech {\bf 0512} (2005) 12012.}
   \ref{CandC}{Cardy,J. and Calabrese,P. \jpa{42}{2009}{504005}.}
   \ref{CaandH}{Casini,H. and Huerta,M. \plb{694}{2010}{167}.}
    \ref{Dow7}{Dowker,J.S. \jpa{25}{1992}{2641}.}
    \ref{Dowcosecs}{Dowker,J.S. {\it On sums of powers of cosecs}, ArXiv:1507.01848.}
    \ref{Jeffery}{Jeffery, H.M. \qjm{6}{1864}{82}.}
   \ref{BMW}{Bueno,P., Myers,R.C. and Witczak--Krempa,W. {\it Universal corner entanglement
   from twist operators} ArXiv:1507.06997.}
  \ref{BMW2}{Bueno,P., Myers,R.C. and Witczak--Krempa,W. {\it Universality of corner
  entanglement in conformal field theories} ArXiv:1505.04804.}
  \ref{BandM}{Bueno,P., Myers,R.C. {\it Universal entanglement for higher dimensional
  cones} ArXiv:1508.00587.}
   \ref{Dowstat}{Dowker,J.S. \jpa{18}{1985}3521.}
   \ref{Dowstring}{Dowker,J.S. {\it Quantum field theory around conical defects} in {\it
   The Formation and Evolution of Cosmic Strings} edited by Gibbons,G.W, Hawking,S.W. and
   Vachaspati,T. (CUP, Cambridge, 1990).}
   \ref{Hung}{Hung,L-Y.,Myers,R.C. and Smolkin,M. {\it JHEP} {\bf 10} (2014) 178.}
\ref{Dow7}{Dowker,J.S. \jpa{25}{1992}{2641}.}
   \ref{B}{Belin,A.,Hung,L-Y., Maloney,A., Matsuura,S., Myers,R.C. and Sierens,T.\break
   {\it JHEP} {\bf 12} (2013) 059.}
   \ref{B2}{Belin,A.,Hung,L-Y.,Maloney,A. and Matsuura,S.
   {\it JHEP01} (2015) 059.}
   \ref{Norlund}{N\"orlund,N.E. \am{43}{1922}{121}.}
    \ref{Norlund1}{N\"orlund,N.E. {\it Differenzenrechnung} (Springer--Verlag, 1924, Berlin.)}
   \ref{Dowconearb}{Dowker,J.S. \prD{36}{1987}{3742}.}
     \ref{Dowren}{Dowker,J.S. \jpamt {46}{2013}{2254}.}
     \ref{DandB}{Dowker,J.S. and Banach,R. \jpa{11}{1978}{2255}.}
     \ref{Dowcen}{Dowker,J.S. {\it Central Differences, Euler numbers and
   symbolic methods} ArXiv: 1305.0500.}
   \ref{Dowcone}{Dowker,J.S. \jpa{10}{1977}{115}.}
   \ref{schulman2}{Schulman,L.S. \jmp{12}{1971}{304}.}
   \ref{DandC}{Dowker,J.S. and Critchley,R. \prD{15}{1977}{1484}.}
     \ref{Thiele}{Thiele,T.N. {\it Interpolationsrechnung} (Teubner, Leipzig, 1909).}
     \ref{Steffensen}{Steffensen,J.F. {\it Interpolation}, (Williams and Wilkins,
    Baltimore, 1927).}
     \ref{Riordan}{Riordan,J. {\it Combinatorial Identities} (Wiley, New York, 1968).}
     \ref{BSSV}{Butzer,P.L., Schmidt,M., Stark,E.L. and Vogt,I. {\it Numer.Funct.Anal.Optim.}
    {\bf 10} (1989) 419.}
      \ref{Dowcascone}{Dowker,J.S. \prD{36}{1987}{3095}.}
      \ref{Stern}{Stern,W. \jram {79}{1875}{67}.}
     \ref{Milgram}{Milgram, M.S., Journ. Maths. (Hindawi) 2013 (2013) 181724.}
     \ref{Perlmutter}{Perlmutter,E. {\it A universal feature of CFT R\'enyi entropy}
     ArXiv:1308.1083 }
     \ref{HMS}{Hung,L.Y., Myers,R.C. and Smolkin,M. {\it Twist operators in
     higher dimensions} ArXiv:1407.6429.}
     \ref{ABD}{Aros,R., Bugini,F. and Diaz,D.E. {\it On the Renyi entropy for
     free conformal fields: holographic and $q$--analog recipes}.ArXiv:1408.1931.}
     \ref{LLPS}{Lee,J., Lewkowicz,A., Perlmutter,E. and Safdi,B.R.{\it R\'enyi entropy.
     stationarity and entanglement of the conformal scalar} ArXiv:1407.7816.}
     \ref{Apps}{Apps,J.S. PhD thesis (University of Manchester, 1996).}
   \ref{CandD}{Candelas,P. and Dowker,J.S. \prD{19}{1979}{2902}.}
    \ref{Hertzberg}{Hertzberg,M.P. \jpa{46}{2013}{015402}.}
     \ref{CaandW}{Callan,C.G. and Wilczek,F. \plb{333}{1994}{55}.}
    \ref{CaandH}{Casini,H. and Huerta,M. \plb{694}{2010}{167}.}
    \ref{Lindelof}{Lindel\"of,E. {\it Le Calcul des Residues} (Gauthier--Villars, Paris,1904).}
    \ref{CaandC}{Calabrese,P. and Cardy,J. {\it J.Stat.Phys.} {\bf 0406} (2004) 002.}
    \ref{MFS}{Metlitski,M.A., Fuertes,C.A. and Sachdev,S. \prB{80}{2009}{115122}.}
    \ref{Gromes}{Gromes, D. \mz{94}{1966}{110}.}
    \ref{Pockels}{Pockels, F. {\it \"Uber die Differentialgleichung $\De
  u+k^2u=0$} (Teubner, Leipzig. 1891).}
   \ref{Diaz}{Diaz,D.E. JHEP {\bf 7} (2008)103.}
  \ref{Minak}{Minakshisundaram,S. {\it J. Ind. Math. Soc.} {\bf 13} (1949) 41.}
    \ref{CaandWe}{Candelas,P. and Weinberg,S. \np{237}{1984}{397}.}
     \ref{Chodos1}{Chodos,A. and Myers,E. \aop{156}{1984}{412}.}
     \ref{ChandD}{Chang,P. and Dowker,J.S. \np{395}{1993}{407}.}
    \ref{LMS}{Lewkowycz,A., Myers,R.C. and Smolkin,M. {\it Observations on
    entanglement entropy in massive QFTs.} ArXiv:1210.6858.}
    \ref{Bierens}{Bierens de Haan,D. {\it Nouvelles tables d'int\'egrales
  d\'efinies}, (P.Engels, Leiden, 1867).}
    \ref{DowGJMS}{Dowker,J.S.  \jpa{44}{2011}{115402}.}
    \ref{Doweven}{Dowker,J.S. {\it Entanglement entropy on even spheres.}
    ArXiv:1009.3854.}
     \ref{Dowodd}{Dowker,J.S. {\it Entanglement entropy on odd spheres.}
     ArXiv:1012.1548.}
    \ref{DeWitt}{DeWitt,B.S. {\it Quantum gravity: the new synthesis} in
    {\it General Relativity} edited by S.W.Hawking and W.Israel (CUP,Cambridge,1979).}
    \ref{Nielsen}{Nielsen,N. {\it Handbuch der Theorie von Gammafunktion}
    (Teubner,Leipzig,1906).}
    \ref{KPSS}{Klebanov,I.R., Pufu,S.S., Sachdev,S. and Safdi,B.R.
    {\it JHEP} 1204 (2012) 074.}
    \ref{KPS2}{Klebanov,I.R., Pufu,S.S. and Safdi,B.R. {\it F-Theorem without
    Supersymmetry} 1105.4598.}
    \ref{KNPS}{Klebanov,I.R., Nishioka,T, Pufu,S.S. and Safdi,B.R. {\it Is Renormalized
     Entanglement Entropy Stationary at RG Fixed Points?} 1207.3360.}
    \ref{Stern}{Stern,W. \jram {79}{1875}{67}.}
    \ref{Gregory}{Gregory, D.F. {\it Examples of the processes of the Differential
    and Integral Calculus} 2nd. Edn (Deighton,Cambridge,1847).}
    \ref{MyandS}{Myers,R.C. and Sinha, A. \prD{82}{2010}{046006}.}
   \ref{RyandT}{Ryu,S. and Takayanagi,T. JHEP {\bf 0608}(2006)045.}
    \ref{Dowcmp}{Dowker,J.S. \cmp{162}{1994}{633}.}
     \ref{Dowjmp}{Dowker,J.S. \jmp{35}{1994}{4989}.}
      \ref{Dowhyp}{Dowker,J.S. \jpa{43}{2010}{445402}.}
       \ref{HandW}{Hertzberg,M.P. and Wilczek,F. \prl{106}{2011}{050404}.}
      \ref{dowkerfp}{Dowker,J.S.\prD{50}{1994}{6369}.}
       \ref{Fursaev}{Fursaev,D.V. \plb{334}{1994}{53}.}
\end{putreferences}

\bye